\documentclass{cjpsuplf}    
\usepackage{epsfig}

\newcommand{\bmm}[1]{\mbox{\boldmath $#1$}}
\newcommand{\btt}[1]{{\bmm #1}_{_T}}
\newcommand{\Pslash}{\kern 0.2 em P\kern -0.56em \raisebox{0.3ex}{/}}
\newcommand{\pslash}{\kern 0.2 em p\kern -0.4em /}
\newcommand{\nslash}{\kern 0.2 em n\kern -0.4em /}
\newcommand{\kslash}{\kern 0.2 em k\kern -0.45em /}
\newcommand{\Sslash}{\kern 0.2 em S\kern -0.56em \raisebox{0.3ex}{/}}
\newcommand{\dslash}{\kern 0.2 em \partial\kern -0.56em \raisebox{0.3ex}{/}}
\newcommand{\xbj}{x_{_B}}
\newcommand{\zh}{z_h}
\def\be{\begin{equation}}
\def\ee{\end{equation}}
\def\bea{\begin{eqnarray}}
\def\ba{\begin{eqnarray}}
\def\eea{\end{eqnarray}}
\def\ea{\end{eqnarray}}
\def\st{{\scriptscriptstyle T}}
\def\slash{\rlap{/}}
\begin{document}
\title{Azimuthal and spin asymmetries in DIS\,\footnote{Lectures
presented at the Advanced Study Institute "Symmetries and Spin"
(Praha-SPIN-2001), Prague}}
\authori{P.J. Mulders}
\addressi{Division of Physics and Astronomy\\
Faculty of Sciences, Vrije Universiteit\\
De Boelelaan 1081, 1081 HV Amsterdam, Netherlands}
\authorii{}     
\addressii{}
\authoriii{}     
\addressiii{}
\headtitle{Azimuthal and spin asymmetries in DIS}
\headauthor{P.J. Mulders}  
\specialhead{P.J. Mulders: Azimuthal and spin asymmetries in DIS}
\evidence{}
\daterec{}    
\suppl{A}  \year{2002}
\setcounter{page}{1}
\maketitle

\begin{abstract}
In these lectures I want to discuss how the structure functions in
deep inelastic scattering relate to quark and gluon correlation functions.
In particular we will consider the issue of intrinsic transverse
momenta of quarks, which becomes important in processes like
1-particle inclusive leptoproduction. Some examples of cross
sections and asymmetries, in particular in polarized scattering
processes are discussed. We also discuss the operator structure
for azimuthal asymmetries and their evolution.
\end{abstract}

\section{Introduction}

The central point of these lectures is the availability of a field
theoretical
framework for the strong interactions and its use to study the
structure of hadrons.
Controlling and selectively probing the nonperturbative regime in high
energy scattering processes is the key to study the structure of hadrons
in the context of QCD. The control parameters for the target and the probe
are the spin and flavor, which in combination with
the kinematical flexibility in scattering processes is used to select the
observable and its gluonic or quarkic nature. Examples are
\begin{tabbing}
aaaaaaaaaaaaaaaa
\= bbbbbbbbbbbbbbbbbbbbbb
\= cccccccccccc \kill
$\ell H  \longrightarrow  \ell^\prime H$
\>(elastic leptoproduction)
\>(spacelike) form factors, \\
$\ell H  \longrightarrow  \ell^\prime X$
\>(inclusive leptoproduction)
\>distribution functions, \\
$\ell H \longrightarrow \ell^\prime h X$
\>(1-particle inclusive
\>distribution and \\
\>\quad leptoproduction)
\> \quad fragmentation functions, \\
$e^+ e^- \longrightarrow h \bar h$
\>(annihilation into $h\bar h$)
\>(timelike) form factors, \\
$e^+ e^- \longrightarrow h X$
\>(1-particle inclusive
\>fragmentation functions, \\
\>\quad annihilation)
\> \\
$H_1 H_2 \longrightarrow \mu^+ \mu^- X$
\>(Drell-Yan scattering)
\>distribution functions.
\end{tabbing}

These notes focus on leptoproduction in which one deals both with
distribution functions (in inclusive processes) and fragmentation
functions (in semi-inclusive processes).

\section{Inclusive Leptoproduction}

\subsection{The hadron tensor}

\begin{figure}[b]
\begin{center}
\begin{minipage}{6 cm}
\epsfig{file=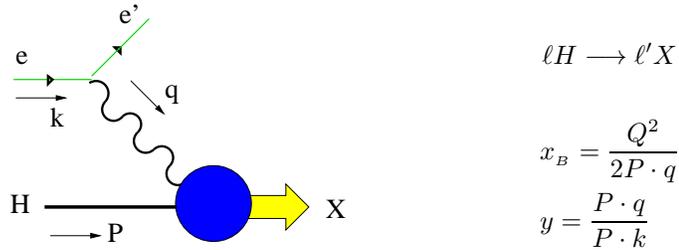,width = 4.5 cm}
\end{minipage}
\begin{minipage}{3 cm}
\begin{eqnarray*}
&&\ell H \longrightarrow \ell^\prime X
\end{eqnarray*}
\begin{eqnarray*}
&&\xbj = \frac{Q^2}{2P\cdot q} \\
&&y = \frac{P\cdot q}{P\cdot k}
\end{eqnarray*}
\end{minipage}
\end{center}
\caption{\label{fig0}
Momenta and invariants in inclusive leptoproduction. The scale is
set by the invariant momentum squared of the virtual photon, $q^2
\equiv -Q^2$, which for a hard process becomes $Q^2 \rightarrow \infty$.}
\end{figure}

For the process $\ell + H \rightarrow \ell^\prime + X$
(see Fig.~\ref{fig0}), the cross section can be separated into a lepton and
hadron part. Although the lepton part is simpler, let us start
with the hadron part,
\bea
&&2M\,W_{\mu\nu}^{(\ell H)}( q; {P S} )
=\frac{1}{2\pi}
\sum_X \int \frac{d^3 P_X}{(2\pi)^3 2P_X^0}
(2\pi)^4 \delta^4 (q + P - P_X)
\nonumber
\\
&& \hspace{5 cm} \times
\langle {P S} |{J_\mu (0)}|P_X \rangle
\langle P_X |{J_\nu (0)}|{P S} \rangle,
\eea
The simplest thing to do is to parametrize this tensor in standard
tensors and structure functions. Instead of the traditional
choice~\cite{Roberts} for
these tensors, $g_{\mu\nu}$ and $P_\mu P_\nu$ and structure functions
$W_1$ and $W_2$, we immediately go to a dimensionless representation,
using a natural space-like momentum (defined by $q$) and a time-like
momentum constructed from $P$ and $q$,
\bea
&&\hat z^\mu = -\hat q^\mu = -\frac{q^\mu}{Q},
\\
&&\hat t^\mu = \frac{\tilde P^\mu}{\sqrt{\tilde P^2}} =
\frac{1}{\sqrt{\tilde P^2}}
\,\left(P^\mu - \frac{P\cdot q}{q^2}\,q^\mu\right)
= \frac{q^\mu + 2\xbj\,P^\mu}{Q}.
\eea
Using hermiticity for the currents, parity invariance and current
conservation one obtains as the most general form the symmetric
tensor
\bea
M\,W_S^{\mu\nu}({q,P}) =
\underbrace{\left\lgroup - g^{\mu\nu} +\hat q^\mu \hat q^\nu
-\hat t^\mu \hat t^\nu\right\rgroup}_{-g_\perp^{\mu\nu}}{F_1}
+ \hat t^\mu\hat t^\nu
\,\underbrace{\left(\frac{F_2}{2\xbj}-{F_1}\right)}_{F_L} ,
\label{param}
\eea
where the structure functions $F_1$ and $F_2$ or the transverse and
longitudinal structure functions, $F_T = F_1$ and $F_L$, depend only
on the for the hadron part relevant invariants $Q^2$ and $\xbj$.
In all equations given here we have omitted target mass effects of
order $M^2/Q^2$.

\subsection{The lepton tensor}

In order to write down the cross section one needs to include the
necessary phase space factors and include the lepton part given by
the tensor
\bea
L_{\mu\nu}^{(\ell H)} (k \lambda ; k^\prime \lambda^\prime)
= 2 k_{\mu}k^\prime_{\nu} + 2 k_{\nu}k^\prime_{\mu}
- Q^2 g_{\mu\nu} +2i\lambda_e\, \epsilon_{\mu\nu\rho\sigma}
q^\rho k^{\sigma} .
\eea
We have included here the (longitudinal) lepton polarization
($\lambda_e = \pm 1$). For later convenience it is useful to rewrite
this tensor also in terms of the space-like and time-like vectors
$\hat q$ and $\hat t$. It is a straightforward exercise to get
\bea
k^\mu = \frac{Q}{2}\,\hat q^\mu + \frac{(2-y) Q}{2y}\,\hat t^\mu
+ \frac{Q\sqrt{1-y}}{y}\,\hat \ell^\mu,
\eea
where $\hat \ell$ is the perpendicular direction defining the lepton
scattering plane (see Fig.~\ref{fig1}). This perpendicular direction
becomes relevant only if other vectors than $P$ and $q$ are present, e.g.
a spin direction in polarized
scattering or the momentum of a produced hadron in 1-particle
inclusive processes. The lepton tensor becomes
\bea
L^{\mu \nu}_{(\ell H)} & = &\frac{Q^2}{y^2} \Biggl[
-2 \left( 1 - y + \frac{1}{2}\,y^2 \right) g_\perp^{\mu \nu}
+ 4(1-y) \hat t^\mu \hat t^\nu
\nonumber \\ && \qquad
+ 4(1-y)\left( \hat \ell^\mu\hat \ell^\nu
+\frac{1}{2}\,g_\perp^{\mu \nu}\right)
+ 2(2 - y)\sqrt{1-y}\,\,\hat t^{\{ \mu}\hat \ell^{\nu \}}
\nonumber \\ &&\qquad
-i\lambda_e\,y(2-y)\,\epsilon_\perp^{\mu \nu}
- 2i\lambda_e\,y\sqrt{1-y}\,\,\hat \ell_\rho \epsilon_\perp^{\rho\,[ \mu}
\hat t^{\nu ]} \Biggr],
\label{leptontensor}
\eea
where $\epsilon_\perp^{\mu\nu} \equiv
\epsilon^{\mu\nu\rho\sigma}\hat t_\rho \hat q_\sigma$.

\subsection{The inclusive cross section}

The cross section for unpolarized lepton and hadron only involves
the first two (symmetric) terms in the lepton tensor and one obtains
\bea
\frac{d\sigma_{OO}}{d\xbj dy} & = & \frac{4\pi\,\alpha^2\,\xbj s}{Q^4}
\Biggl\{ \left( 1-y + \frac{1}{2}\,y^2\right) {F_T(\xbj,Q^2)}
+ \left( 1 -y\right) {F_L(\xbj,Q^2)}\Biggr\}.
\eea
As soon as the exchange of a $Z^0$ boson becomes important the
hadron tensor is no longer constrained by parity invariance and a
third structure function $F_3$ becomes important.

\subsection{Target polarization}

The use of polarization in leptoproduction provides new ways to probe
the hadron target.
For a spin 1/2 particle the initial state is described by a
2-dimensional spin density matrix $\rho = \sum_\alpha \vert \alpha\rangle
p_\alpha \langle \alpha\vert$ describing the probabilities $p_\alpha$
for a variety of spin possibilities. This density matrix is hermitean
with Tr$\,\rho$ = 1. It can in the target rest frame be expanded
\be
\rho_{ss^\prime} = \frac{1}{2}
\left( 1 + \bmm S\cdot \bmm \sigma_{ss^\prime}\right),
\ee
where $\bmm S$ is the spin vector. When
$\vert \bmm S\vert = 1$ one has a pure state (only one state $\vert
\alpha\rangle$ and $\rho^2 = \rho$), when $\vert \bmm S\vert \le 1$
one has an ensemble of states. For the
case $\vert \bmm S\vert = 0$ one has simply an averaging over spins,
corresponding to an unpolarized ensemble. To include spin one
could generalize the hadron tensor to a matrix in spin space,
$\tilde W_{s^\prime s}^{\mu\nu}(q,P)$ $\propto$
$\langle P, s^\prime\vert J^\mu\vert X\rangle\langle X\vert
J^\nu\vert P, s>$
depending only on the momenta or one can look at the tensor
$\sum_\alpha p_\alpha \tilde W_{\alpha\alpha}^{\mu\nu}(q,P)$,
which is given by
\be
W^{\mu\nu}(q,P,S) = \mbox{Tr}\left( \rho(P,S) \tilde W^{\mu\nu}(q,P)
\right),
\label{defS}
\ee
with the spacelike spin vector $S$ appearing {\em linearly} and
in an arbitrary frame satisfying $P\cdot S = 0$. It has
invariant length $-1 \le S^2 \le 0$.
It is convenient to write the spin vector as
\bea
S^\mu = \frac{\lambda}{M}\left( P^\mu - \frac{2\xbj
M^2}{Q^2}\,q^\mu\right) + S_\perp^\mu,
\label{inclspin}
\eea
with $\lambda = M(S\cdot q)/(P\cdot q)$. For a pure state one
has $\lambda^2 + \bmm S_\perp^2 = 1$.
Using symmetry constraints one obtains
for electromagnetic interactions (parity conservation)
an antisymmetric part in the hadron tensor,
\bea
M\,W_A^{\mu\nu}({q,P,S}) =
\underbrace{-i\,{\lambda}\,
\frac{\epsilon^{\mu\nu\rho\sigma}P_\rho q_\sigma}
{P\cdot q}}_{-i\,{\lambda}\,\epsilon_\perp^{\mu\nu}}
\,{g_1}
+ i\,\frac{2M\xbj}{Q}
\,\hat t_{\mbox{}}^{\,[\mu}\epsilon_\perp^{\nu ]\rho} {S_{\perp\rho}}
\,{g_T}.
\label{wanti}
\eea
The polarized part of the cross section becomes
\bea
\frac{d\sigma_{LL}}{d\xbj dy} & = & \lambda_e\,\frac{4\pi\,\alpha^2}{Q^2}
\Biggl\{ {\lambda}\,\left( 1-\frac{y}{2} \right){g_1(\xbj,Q^2)}\nonumber
\\ && \qquad\qquad
-\vert {S_\perp}\vert\,\cos\,\phi_S^\ell\,\frac{2M\xbj}{Q} \sqrt{1-y}
\,\,{g_T(\xbj,Q^2)}\Biggr\}.
\eea

\section{Semi-inclusive leptoproduction}

\subsection{The hadron tensor}

\begin{figure}[t]
\begin{center}
\begin{minipage}{8.5cm}
\epsfig{file=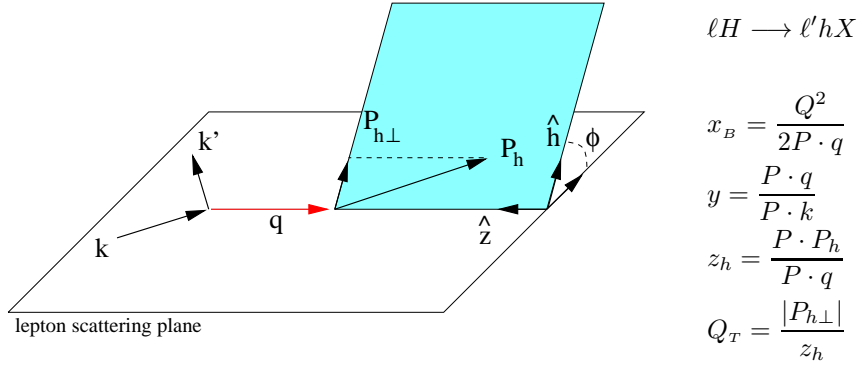,width=8.5cm}
\end{minipage}
\begin{minipage}{2.5 cm}
\begin{eqnarray*}
&&\ell H \longrightarrow \ell^\prime h X
\end{eqnarray*}
\begin{eqnarray*}
&&\xbj = \frac{Q^2}{2P\cdot q} \\
&&y = \frac{P\cdot q}{P\cdot k}\\
&&z_h = \frac{P\cdot P_h}{P\cdot q}\\
&&Q_\st = \frac{\vert P_{h\perp}\vert}{z_h}
\end{eqnarray*}
\end{minipage}
\end{center}
\caption{\label{fig1}
Kinematics for 1-particle inclusive leptoproduction.}
\end{figure}

More flexibility in probing
new aspects of hadron structure is achieved in semi-inclusive
scattering processes. For instance in 1-particle inclusive
measurements one can measure azimuthal dependences in the cross
sections. The central
object of interest for 1-particle
inclusive leptoproduction, the hadron tensor, is given by
\bea
&&2M{\cal W}_{\mu\nu}^{(\ell H)}( q; {P S; P_h S_h} )
=
\sum_X \int \frac{d^3 P_X}{(2\pi)^3 2P_X^0}
\delta^4 (q + P - P_X - P_h)
\nonumber
\\
&& \hspace{2.5 cm} \times
\langle {P S} |{J_\mu (0)}|P_X; {P_h S_h} \rangle
\langle P_X; {P_h S_h} |{J_\nu (0)}|{P S} \rangle,
\eea
where $P,\ S$ and $P_h,\ S_h$ are the momenta and spin vectors
of target hadron and produced hadron,
$q$ is the (space-like) momentum transfer with $-q^2$ = $Q^2$ sufficiently
large.
The kinematics is illustrated in Fig.~\ref{fig1}, where also the scaling
variables are introduced.
For the parametrization of the hadron tensor in terms of structure
functions it is useful to introduce the directions $\hat q$ and
$\hat t$ as before and using the vector $P_h$ to construct
a vector that is orthogonal to these vectors. For the situation that
$P\cdot P_h$ is ${\cal O}(Q^2)$ (current fragmentation!) one
finds that
\bea
q_\st^\mu = q^\mu +\xbj\,P^\mu - \frac{P_{h}^\mu}{\zh} =
-\frac{P_{h\perp}^\mu}{\zh} \equiv -Q_T\,\hat h^\mu ,
\eea
is such a vector.
This vector is proportional to the transverse momentum of the
outgoing hadron with respect ot $P$ and $q$. It can also be
considered as
the transverse momentum of the photon with respect to the hadron
momenta $P$ and $P_h$.
For an unpolarized (or spin 0) hadron in the final state
the symmetric part of the tensor is given by
\bea
M{\cal W}_S^{\mu\nu}({q,P,P_h}) &=&
- g_\perp^{\mu\nu}\,{\cal H}_T
+ \hat t^\mu\hat t^\nu\,{\cal H}_L
\nonumber \\ &&
+ \hat t^{\,\{\mu}\hat h^{\nu\}}\,{\cal H}_{LT}
+ \left\lgroup 2\,\hat h^\mu \hat h^\nu + g_\perp^{\mu\nu}\right\rgroup
{\cal H}_{TT} .
\eea
Noteworthy is that also an antisymmetric term in the tensor is allowed,
\bea
M{\cal W}_A^{\mu\nu}({q,P,P_h}) =
- i\hat t^{\,[\mu}\hat h^{\nu]}\,{\cal H}^\prime_{LT}.
\label{sidiswanti}
\eea

\subsection{The semi-inclusive cross section}

Clearly the lepton tensor in Eq.~\ref{leptontensor} is able to distinguish
all the structures in the semi-inclusive hadron tensor. The
symmetric part gives the cross section for unpolarized leptons,
\bea
\frac{d\sigma_{OO}}{d\xbj dy\,d\zh d^2q_\st}
& = & \frac{4\pi\,\alpha^2\,s}{Q^4}\,\xbj \zh
\Biggl\{ \left( 1-y+\frac{1}{2}\,y^2 \right){\cal H}_T
+ (1-y)\,{\cal H}_L
\nonumber\\ && \qquad \qquad \qquad
\mbox{} - (2-y)\sqrt{1-y}\,\cos \phi_h^\ell\,\,{\cal H}_{LT}
\nonumber\\ && \qquad \qquad \qquad
\mbox{} + (1-y)\,\cos 2\phi_h^\ell\,\,{\cal H}_{TT}
\Biggr\}
\eea
while the antisymmetric part gives
the cross section for a polarized lepton
(note the target is not polarized!)
\bea
\frac{d\sigma_{LO}}{d\xbj dy\,d\zh d^2q_\st}
= \lambda_e\,\frac{4\pi\,\alpha^2}{Q^2}\,\zh
\,\sqrt{1-y}\,\sin \phi_h^\ell\,\,{\cal H}^\prime_{LT} .
\eea
Of course many more structure functions appear for polarized targets or
if one considers polarimetry in the final state.

\section{Quark correlation functions in leptoproduction}

\begin{figure}[t]
\begin{center}
\epsfig{file=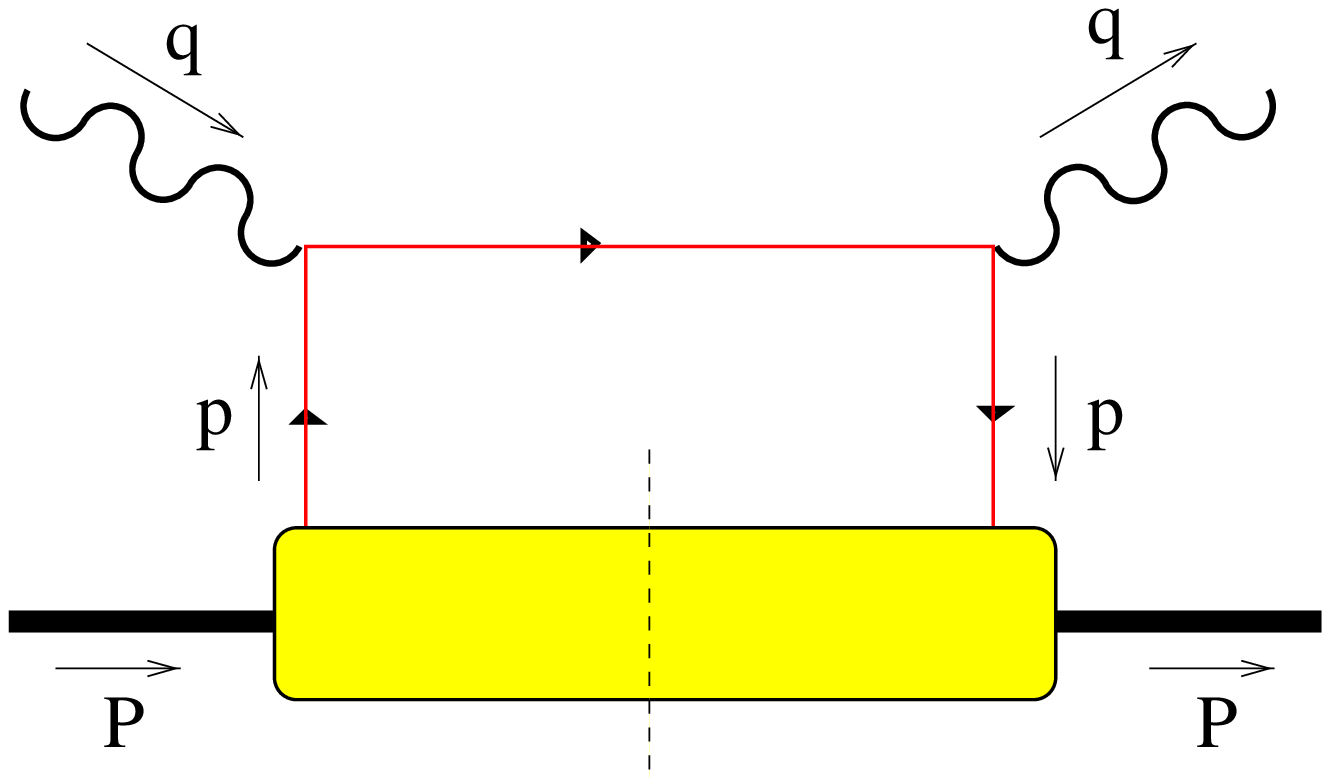,width=4.5cm}
\hspace{2 cm}
\epsfig{file=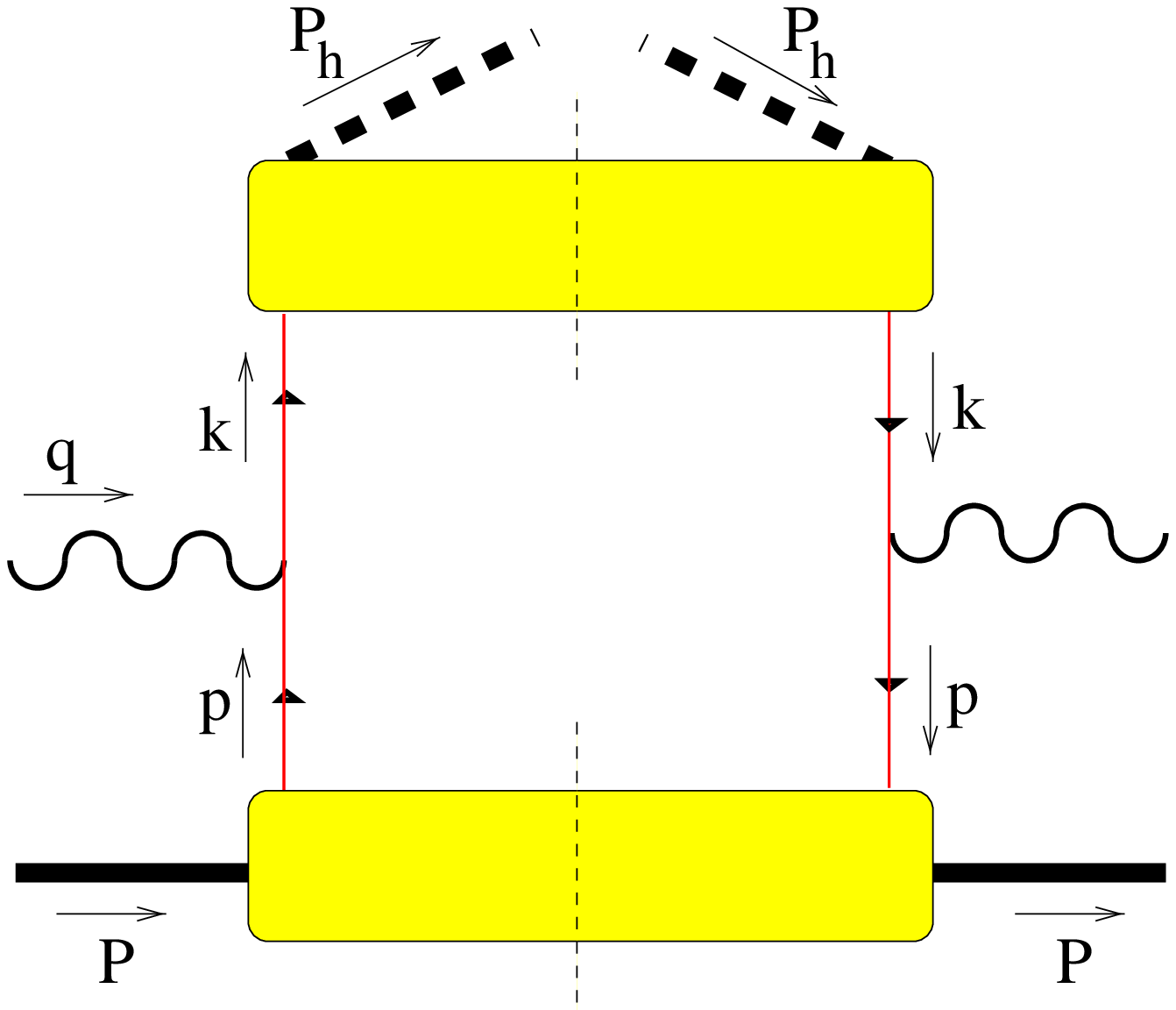,width=4.5cm}
\end{center}
\caption{\label{fig2}
The simplest (parton-level) diagrams representing the squared amplitude
in lepton hadron inclusive scattering (left) and semi-inclusive scattering
(right). In both cases also the diagram with opposite fermion flow has to
be added.}
\end{figure}

Within the framework of QCD and knowing that the photon or $Z^0$ current
couples to the quarks, it is possible to write down a diagrammatic expansion
for leptoproduction, with in the deep inelastic limit
($Q^2 \rightarrow \infty$) as relevant diagrams only the
ones given in Fig.~\ref{fig2} for
inclusive and 1-particle inclusive scattering respectively.
The expression for ${\cal W}_{\mu\nu}$ can be rewritten as a nonlocal
product of currents and
it is a straightforward exercise to show
by inserting the currents
$j_\mu(x) = :\overline \psi(x) \gamma_\mu \psi(x):$
that for 1-particle inclusive scattering
one obtains in tree approximation
\bea
&&2M{\cal W}_{\mu\nu}( q; P S; P_h S_h ) =
\nonumber \\ & & \qquad \mbox{} =
\frac{1}{(2\pi)^4} \int d^4x\ e^{iq\cdot x}\,
\langle P S |:\overline \psi_j (x) (\gamma_\mu)_{jk} \psi_k(x):
\sum_X |X; P_h S_h \rangle \nonumber \\
&&\qquad \qquad \qquad \qquad \qquad \times \langle X; P_h S_h |
:\overline \psi_l(0) (\gamma_\nu)_{li} \psi_i(0):|P S \rangle \nonumber \\
& & \qquad \mbox{}
= \frac{1}{(2\pi)^4} \int d^4x\ e^{iq\cdot x}\,
\langle P S |\overline \psi_j (x) \psi_i(0) \vert P S \rangle
(\gamma_\mu)_{jk}
\nonumber \\
&& \qquad \qquad \qquad  \langle 0 \vert \psi_k(x)
\sum_X |X; P_h S_h \rangle
\langle X; P_h S_h | \overline \psi_l(0) \vert 0 \rangle (\gamma_\nu)_{li}
\nonumber \\
&& \qquad \quad \mbox{}
+ \frac{1}{(2\pi)^4} \int d^4x\ e^{iq\cdot x}\,
\langle P S |\psi_k (x) \overline \psi_l(0) \vert P S \rangle
(\gamma_\nu)_{li} \nonumber \\
&&\qquad \qquad \qquad  \langle 0 \vert \overline \psi_j(x)  \sum_X |X; P_h
S_h \rangle \langle X; P_h S_h |
\psi_i(0) \vert 0 \rangle (\gamma_\mu)_{jk}, \nonumber \\
& & \qquad \mbox{} =
\int d^4p \,d^4k\,\delta^4(p+q-k) \ \mbox{Tr}\left( \Phi(p) \gamma_\mu
\Delta(k) \gamma_\nu \right)
+ \left\{ \begin{array}{c} q \leftrightarrow -q \\ \mu \leftrightarrow \nu
\end{array} \right\},
\label{basic}
\eea
where
\bea
&& \Phi_{ij}(p) =  \frac{1}{(2\pi)^4} \int d^4\xi\ e^{ip\cdot \xi}\,
\langle P S |\overline \psi_j (0) \psi_i(\xi)|P S \rangle, \nonumber \\
&& \Delta_{kl}(k) = \frac{1}{(2\pi)^4} \int d^4\xi\ e^{ik\cdot \xi}\,
\langle 0 \vert \psi_k(\xi) \sum_X |X; P_h S_h \rangle \langle X; P_h S_h |
\overline \psi_l(0) \vert 0 \rangle. \nonumber
\eea
Note that in $\Phi$ (quark
production) a summation over colors is assumed, while in $\Delta$
(quark decay)
an averaging over colors is assumed. The quantities $\Phi$ and $\Delta$
correspond to the blobs in Fig.~\ref{fig2} and parametrize the soft physics.
Soft refers to all invariants of momenta being small as compared to the hard
scale, i.e. for $\Phi(p)$ one has $p^2 \sim p\cdot P \sim P^2 = M^2 \ll Q^2$.

In general many more diagrams have to be considered in evaluating the
hadron tensors, but in the deep inelastic limit they can be neglected
or considered as corrections to the soft blobs. We return to this later.

As mentioned above,
the relevant structural information for the
hadrons is contained in soft parts (the blobs in Fig.~\ref{fig2})
which represent specific matrix elements of quark fields.
The form of $\Phi$ is constrained by hermiticity, parity and time-reversal
invariance. The quantity depends besides the quark momentum $p$ on the
target momentum $P$ and the spin vector $S$ and one must have
\begin{eqnarray}
&& \mbox{[Hermiticity]} \quad \Rightarrow \quad
\Phi^\dagger (p,P,S) = \gamma_0 \,\Phi(p,P,S)\,\gamma_0 ,
\\
&& \mbox{[Parity]} \quad \Rightarrow \quad
\Phi(p,P,S) = \gamma_0 \,\Phi(\bar p,\bar P,-\bar S)\,\gamma_0 ,
\\
&&\mbox{[Time\ reversal]} \ \Rightarrow
\ \Phi^\ast(p,P,S) = (-i\gamma_5 C)\,\Phi(\bar p,\bar P,
\bar S)\,(-i\gamma_5 C) ,
\end{eqnarray}
where $C$ = $i\gamma^2 \gamma_0$, $-i\gamma_5 C$= $i\gamma^1\gamma^3$
and $\bar p$ = $(p^0,-\bmm p)$.
The most general way to parametrize $\Phi$ using only the constraints
from hermiticity and parity invariance, is~\cite{RS79,TM95}
\bea
\Phi(p,P,S) & = &
M\,A_1 + A_2\,\slash P + A_3 \slash p
+ i\,A_4\,\frac{[\slash P,\slash p]}{2M}
\nonumber \\ & &
+ i\,A_{5}\,(p\cdot S) \gamma_5
+ M\,A_6 \,\slash S \gamma_5
+ A_7\,\frac{(p\cdot S)}{M}\,\slash P \gamma_5
\nonumber \\ & &
+ A_8\,\frac{(p\cdot S)}{M}\,\slash p \gamma_5
+ A_9\,\frac{[\slash P,\slash S]}{2}\,\gamma_5
+ A_{10}\,\frac{[\slash p,\slash S]}{2}\,\gamma_5
\nonumber \\ & &
+ A_{11}\,\frac{(p\cdot S)}{M}\,\frac{[\slash P,\slash p]}{2M}\,\gamma_5
+ A_{12}\,\frac{\epsilon_{\mu \nu \rho \sigma}\gamma^\mu P^\nu
p^\rho S^\sigma}{M},
\label{lorentz}
\eea
where the first four terms do not involve
the hadron polarization vector. Hermiticity requires all the amplitudes
$A_i$ = $A_i(p\cdot P, p^2)$ to be real. The amplitudes $A_4$, $A_5$
and $A_{12}$ vanish when also time reversal invariance applies.

\section{Inclusive scattering}

\subsection{The relevant soft parts}

In order to find out which information in the soft parts
is important in a hard process one needs to realize
that the hard scale $Q$ leads in a natural way to the use of lightlike
vectors $n_+$ and $n_-$ satisfying $n_+^2 = n_-^2 = 0$ and $n_+\cdot n_-$
= 1. For inclusive scattering one parametrizes the momenta
\[
\left.
\begin{array}{l} q^2 = -Q^2 \\
P^2 = M^2\\
2\,P\cdot q = \frac{Q^2}{\xbj} \\
\end{array} \right\}
\longleftrightarrow \left\{
\begin{array}{l}
q =\ \frac{Q}{\sqrt{2}}\,n_- \ - \ \frac{Q}{\sqrt{2}}\,n_+
\\ \mbox{} \\
P = \frac{\xbj M^2}{Q\sqrt{2}}\,n_-
+ \frac{Q}{\xbj \sqrt{2}}\,n_+
\end{array}
\right.
\]
The above are the external momenta. Next turn to the internal
momenta, looking at the left diagram in Fig.~\ref{fig2}.
In the soft part actually {\em all} momenta, that is $p$ {\em and} $P$
have a minus
component that can be neglected compared to that in the hard part,
since otherwise $p\cdot P$ would be hard. Thus because $p$ must have
only a hard plus component, $q$ has two hard components and $k$ being
the current jet also must be soft, i.e. only can have one large
lightcone component, one must have
\begin{eqnarray*}
p &=& \quad\ldots\quad + \frac{Q}{\sqrt{2}}\,n_+ ,
\\
q &=& \frac{Q}{\sqrt{2}}\,n_- - \frac{Q}{\sqrt{2}}\,n_+ ,
\\
p+q = k &=& \frac{Q}{\sqrt{2}}\,n_- + \quad\ldots\ .
\end{eqnarray*}
where the \ldots parts indicate (negligible) $1/Q$ terms.

Also the transverse component is not relevant for the hard part.
One thus sees that for inclusive scattering the only relevant
dependence of the soft part is the $p^+$ dependence. Moreover,
the above requirements on the internal momenta already indicate
that the lightcone fraction $x = p^+/P^+$ must be equal to $\xbj$.
This will come out when we do the actual calculation in one
of the next sections.

The minus component $p^- \equiv p\cdot n_+$ and transverse components
thus can be integrated over restricting the nonlocality in $\Phi(p)$.
The relevant soft part then is some Dirac trace
of the quantity~\cite{Soper77,Jaffe83}
\bea
\Phi_{ij}(x) & = &
\int dp^-\,d^2p_\st \ \Phi_{ij}(p,P,S)
\nonumber \\
& = &
\left. \int \frac{d\xi^-}{2\pi}\ e^{ip\cdot \xi}
\,\langle P,S\vert \overline \psi_j(0) \psi_i(\xi)
\vert P,S\rangle \right|_{\xi^+ = \xi_\st = 0},
\eea
depending on the lightcone fraction $x = p^+/P^+$.
To be precise one puts in the full form for the quark momentum,
\be
p = x\,P^+n_+ + \frac{p^2 + \bmm p_\st^2}{2x\,P^+}\,n_- + p_\st,
\label{quarkmom}
\ee
and performs the integration over $\Phi(p)$ using
\be
\int dp^-\,d^2p_\st \ldots = \frac{\pi}{P^+}\int d(p\cdot P)\,dp^2
\ldots \ .
\ee
When one wants to calculate the leading order in $1/Q$ for a hard
process, one only needs to look at leading parts in $M/P^+$ because
$P^+ \propto Q$ (see opening paragraph of this section)~\cite{JJ92}.
In this case that turns out
to be the part proportional to $(M/P^+)^0$,
\be
\Phi(x) =
\frac{1}{2}\,\Biggl\{
f_1(x)\,\nslash_+
+ \lambda\,g_1(x)\, \gamma_5\,\nslash_+
+ h_1(x)\,\frac{\gamma_5\,[\Sslash_\perp,\nslash_+]}{2}\Biggr\}
+ {\cal O}\left(\frac{M}{P^+}\right)
\ee
The precise expression of the functions $f_1(x)$, etc. as integrals
over the amplitudes can be easily written down.

\subsection{Calculating the inclusive cross section}

Using field theoretical methods the left diagram
in Fig.~\ref{fig2} can now be calculated.
Omitting the sum over flavors ($\sum_a$),
the quark charges $e_a^2$ and the
$(q\leftrightarrow -q, \mu\leftrightarrow \nu)$ 'antiquark' diagram,
the symmetric part of the hadron tensor the result is
\be
2M\,W^{\mu \nu}(P,q)
= \int dp^-\,dp^+\,d^2p_\perp \ \mbox{Tr}\left(\Phi(p)
\,\gamma^\mu \Delta(p+q) \gamma^\nu\right),
\ee
where
\be
\Delta(k) = (\slash k + m)\,\delta(k^2-m^2)
\approx \frac{\slash n_-}{2}\,\delta(k^+) ,
\ee
and in the approximation anything proportional to $1/Q^2$ has
been neglected. One obtains
\begin{eqnarray}
2M\,W_S^{\mu \nu}(P,q)
&=& \int dp^-\,dp^+\,d^2p_\perp \ \frac{1}{2}\,\mbox{Tr}\left(\Phi(p)
\,\gamma^\mu \,\gamma^+\,\gamma^\nu\right)
\, \delta ( p^+ + q^+) \nonumber \\
&= & -g_\perp^{\mu\nu}\ \left. \mbox{Tr}\left(\gamma^+\,\Phi(x) \right)
\right|_{x = \xbj} \quad
= \quad -g_\perp^{\mu\nu}\,f_1(\xbj).
\label{inclcalc}
\end{eqnarray}
Antiquarks arise from the diagram with opposite fermion flow,
proportional to
$\mbox{Tr}\left(\overline\Phi(p)
\,\gamma^\nu \overline\Delta(k) \gamma^\mu\right)$
with
\be
\overline \Phi_{ij}(p) =
\frac{1}{(2\pi)^4} \int d^4\xi\ e^{-ip\cdot \xi}
\,\langle P S |\psi_i (\xi) \overline \psi_j(0)|P S \rangle .
\ee
The {\em proper} definition of antiquark distributions starts from
$\Phi^c(x)$ containing antiquark distributions $\bar f_1(x)$, etc.
The quantity $\Phi^c(p)$ is obtained from $\Phi(p)$ after
the replacement of $\psi$ by $\psi^C = C\overline\psi^T$.
One then finds $\overline \Phi(p)$ = $-C(\Phi^c)^TC^\dagger$, i.e.
one has to be aware of sign differences.
Symmetry relations between quark and antiquark relations
can be obtained using the anticommutation relations for fermions,
giving $\overline \Phi_{ij}(p) = - \Phi_{ij}(-p)$. One finds that
$\bar f_1(x) = -f_1(-x)$, $\bar g_1(x) = g_1(-x)$, and
$\bar h_1(x) = -h_1(-x)$.
Finally, after including the flavor summation and the quark
charges squared one can compare the result with
Eq.~\ref{param} to obtain for the structure function
\be
2F_1(\xbj) = \sum_a e_a^2\,\left(f_1^a(\xbj) + f_1^{\bar a}(\xbj)\right),
\ee
while $F_L(\xbj) = 0$ (Callan-Gross relation).

The antisymmetric part of $W^{\mu\nu}$ in the above calculation is
left as an exercise. The answer is
\begin{eqnarray}
2M\,W_A^{\mu \nu}(P,q)
&=& i\,\epsilon_\perp^{\mu\nu}\,g_1(\xbj),
\end{eqnarray}
which after inclusion of antiquarks, flavor summation gives
(cf Eq.~\ref{wanti})
\be
2g_1(\xbj) = \sum_a e_a^2\,\left(g_1^a(\xbj) + g_1^{\bar a}(\xbj)\right).
\ee

\subsection{Interpretation of the functions}

The functions $f_1$, $g_1$ and $h_1$ can be obtained from the
correlator $\Phi(x)$ after tracing with the appropriate
Dirac matrix,
\bea
f_1(x) & = &
\left. \int \frac{d\xi^-}{4\pi}\ e^{ip\cdot \xi}
\,\langle P,S\vert \overline \psi(0) \gamma^+ \psi(\xi)
\vert P,S\rangle \right|_{\xi^+ = \xi_\st = 0},
\\
\lambda\,g_1(x) & = &
\left. \int \frac{d\xi^-}{4\pi}\ e^{ip\cdot \xi}
\,\langle P,S\vert \overline \psi(0) \gamma^+\gamma_5 \psi(\xi)
\vert P,S\rangle \right|_{\xi^+ = \xi_\st = 0},
\\
S_\st^i\,h_1(x) & = &
\left. \int \frac{d\xi^-}{4\pi}\ e^{ip\cdot \xi}
\,\langle P,S\vert \overline \psi(0) \,i\sigma^{i+}\gamma_5\,\psi(\xi)
\vert P,S\rangle \right|_{\xi^+ = \xi_\st = 0},
\eea
By introducing {\em good} and {\em bad} fields
$\psi_\pm \equiv \frac{1}{2}\gamma^\mp\gamma^\pm \psi$, one
sees that $f_1$ can be rewritten as
\bea
f_1(x) & = &
\left. \int \frac{d\xi^-}{2\pi\sqrt{2}}\ e^{ip\cdot \xi}
\,\langle P,S\vert \psi^\dagger_+(0) \psi_+(\xi)
\vert P,S\rangle \right|_{\xi^+ = \xi_\st = 0}
\nonumber \\
& = & \frac{1}{\sqrt{2}}\sum_n \left| \langle P_n\vert
\psi_+\vert P\rangle\right|^2\,\delta\left(P_n^+ - (1-x)P^+\right) ,
\eea
i.e. it is a quark lightcone momentum distribution. For the functions
$g_1$ and $h_1$ one needs in addition the projectors on quark chirality
states, $P_{R/L} = \frac{1}{2}(1\pm \gamma_5)$, and on quark transverse
spin states~\cite{Artru,JJ92},
$P_{\uparrow/\downarrow} = \frac{1}{2}(1\pm \gamma^i
\gamma_5)$ to see that
\bea
&&
f_1(x) = f_{1R}(x) + f_{1L}(x) = f_{1\uparrow}(x) + f_{1\downarrow}(x),
\\ &&
g_1(x) = f_{1R}(x) - f_{1L}(x),
\\ &&
h_1(x) = f_{1\uparrow}(x) - f_{1\downarrow}(x).
\eea
One sees some trivial bounds such as $f_1(x) \ge 0$ and
$\vert g_1(x)\vert \le f_1(x)$.
Since $P_n^+ \le 0$ and sees $x \le 1$. From the antiquark
distribution $\bar f_1(x)$ and its relation to $f_1(x)$ one
obtains $x \ge -1$, thus the support of the functions is
$-1 \le x \le 1$.

\subsection{Bounds on the distribution functions}

The trivial bounds on the distribution functions
($\vert h_1(x)\vert \le f_1(x)$ and $\vert g_1(x)\vert
\le f_1(x)$) can be sharpened. For instance
one can look explicitly at the structure in Dirac space of the
correlation function $\Phi_{ij}$. Actually, we will look at the
correlation functions $(\Phi\,\gamma_0)_{ij}$, which involves
at leading order matrix elements $\psi_{+j}^\dagger (0)\psi_{+i}(\xi)$.
One has in Weyl representation
($\gamma^0 = \rho^1$,
$\gamma^i = -i\rho^2\sigma^i$,
$\gamma_5 = i\gamma^0\gamma^1\gamma^2\gamma^3 = \rho^3$)
the matrices
\begin{eqnarray*}
P_+ =
\left\lgroup \begin{array}{rrrr}
1 & 0 & 0 & 0 \\
0 & 0 & 0 & 0 \\
0 & 0 & 0 & 0 \\
0 & 0 & 0 & 1
\end{array}\right\rgroup,
P_+\gamma_5 =
\left\lgroup \begin{array}{rrrr}
1 & 0 & 0 & 0 \\
0 & 0 & 0 & 0 \\
0 & 0 & 0 & 0 \\
0 & 0 & 0 & -1
\end{array}\right\rgroup,
P_+\gamma^1\gamma_5 =
\left\lgroup \begin{array}{rrrr}
0 & 0 & 0 & 1 \\
0 & 0 & 0 & 0 \\
0 & 0 & 0 & 0 \\
1 & 0 & 0 & 0
\end{array}\right\rgroup .
\end{eqnarray*}
The good projector only leaves two (independent) Dirac spinors, one
righthanded (R), one lefthanded (L).
On this basis of good R and L spinors the for hard scattering processes
relevant matrix $(\Phi\slash n_-)$ is given by
\be
(\Phi\,\slash n_-)_{ij}(x) =
\left\lgroup \begin{array}{cc}
f_1 + \lambda\,g_1 &  (S_\st^1-i\,S_\st^2)\,h_1 \\
& \\
(S_\st^1+i\,S_\st^2)\,h_1 & f_1 - \lambda\,g_1
\end{array}\right\rgroup
\ee
One can also turn the $S$-dependent correlation function $\Phi$
defined in analogy with $W(q,P,S)$ in Eq.~\ref{defS}
into a matrix in the nucleon spin space. If
\bea
\Phi (x; P,S) &=& \Phi_O + \lambda\,\Phi_L + S_\st^1\,\Phi_\st^1
+ S_\st^2\,\Phi_\st^2 ,
\eea
then one has on the basis of spin 1/2 target states with $\lambda = +1$
and $\lambda = -1$ respectively
\be
\Phi_{ss^\prime}(x) =
\left\lgroup \begin{array}{cc}
\Phi_O + \Phi_L & \Phi_\st^1 - i\,\Phi_\st^2 \\
& \\
\Phi_\st^1 + i\,\Phi_\st^2 & \Phi_O - \Phi_L \\
\end{array}\right\rgroup
\ee
The matrix relevant for bounds is the matrix $M$ = $(\Phi \slash n_-)^T$
(for this matrix one has $v^\dagger M v \ge 0$ for any direction $v$).
On the basis $+R$, $-R$, $+L$ and $-L$ it becomes
\bea
(\Phi(x)\,\slash n_-)^T &=&
\left\lgroup \begin{array}{cccc}
f_1 + g_1 & 0 & 0 & 2\,h_1 \\
& & &\\
0 & f_1 - g_1 & 0 & 0 \\
& & &\\
0 & 0 & f_1 - g_1 & 0 \\
& & &\\
2\,h_1 & 0 & 0 & f_1 + g_1
\end{array}\right\rgroup
\ \begin{array}{c}
\includegraphics[width = 0.8 cm]{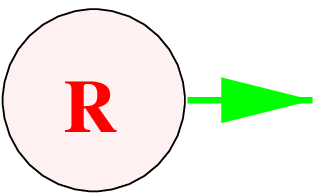}\\[0.3cm]
\includegraphics[width = 0.8 cm]{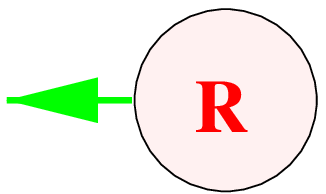}\\[0.3cm]
\includegraphics[width = 0.8 cm]{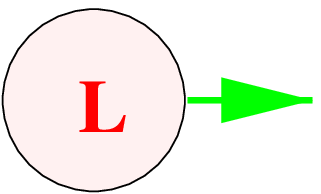}\\[0.3cm]
\includegraphics[width = 0.8 cm]{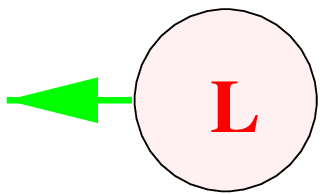}
\end{array}
\label{prod1}
\\
&&\mbox{}\hspace{0.7cm}
\includegraphics[width = 0.8 cm]{helrr.eps}
\hspace{0.7cm}\includegraphics[width = 0.8 cm]{hellr.eps}
\hspace{1.0cm}\includegraphics[width = 0.8 cm]{helrl.eps}
\hspace{0.7cm}\includegraphics[width = 0.8 cm]{helll.eps}
\nonumber
\eea
Of this matrix any diagonal matrix element must always be positive,
hence the eigenvalues must be positive, which  gives a bound on the
distribution functions stronger than the trivial bounds, namely
\be
\vert h_1(x)\vert \le \frac{1}{2}\left( f_1(x) + g_1(x)\right)
\ee
known as the Soffer bound~\cite{Soffer}.

\subsection{Sum rules}

For the functions appearing in the soft parts, and thus also for the
structure functions, one can derive sum rules. Starting with
the traces defining the quark distributions,
\begin{eqnarray*}
f_1(x) & = &
\left. \int \frac{d\xi^-}{4\pi}\ e^{ip\cdot \xi}
\,\langle P,S\vert \overline \psi(0) \gamma^+ \psi(\xi)
\vert P,S\rangle \right|_{\xi^+ = \xi_\st = 0},
\\
g_1(x) & = &
\left. \int \frac{d\xi^-}{4\pi}\ e^{ip\cdot \xi}
\,\langle P,S\vert \overline \psi(0) \gamma^+\gamma_5 \psi(\xi)
\vert P,S\rangle \right|_{\xi^+ = \xi_\st = 0},
\end{eqnarray*}
and integrating over $x = p^+/P^+$ one obtains (using symmetry relation
as indicated above to eliminate antiquarks $\bar f_1$),
\bea
\int_0^1 dx\,\left( f_1(x) - \bar f_1(x)\right)
= \int_{-1}^1 dx\ f_1(x)
= \frac{\langle P,S\vert \overline \psi(0) \gamma^+ \psi(0)
\vert P,S\rangle }{2P^+},
\eea
which as we have seen in the section on elastic scattering is nothing
else than a form factor at zero momentum transfer, i.e. the number
of quarks of that particular flavor. Similarly one finds the sum rule
\bea
\int_0^1 dx\,\left( g_1(x) + \bar g_1(x)\right)
= \int_{-1}^1 dx\ g_1(x)
= \frac{\langle P,S\vert \overline \psi(0) \gamma^+\gamma_5 \psi(0)
\vert P,S\rangle }{2P^+},
\label{axsumr}
\eea
which precisely is the axial charge $g_A$ for a particular quark flavor.
These relations of quark distributions and matrix elements
underly sum rules for
the structure functions, such as the Bjorken sum rule
\be
\int_0^1 d\xbj\ \left(g_1^p(\xbj,Q^2)-g_1^n(\xbj,Q^2)\right)
= \frac{1}{6}\left(g_A^u - g_A^d\right)
= \frac{1}{6}\,G_A^{n\rightarrow p}(0),
\ee
relating the polarized structure function to the axial charge measured
in neutron $\beta$-decay which also can be expressed in quark axial charges.

\section{1-particle inclusive scattering}

\subsection{The relevant distribution functions}

For 1-particle inclusive scattering one parametrizes the momenta
\[
\left.
\begin{array}{l} q^2 = -Q^2 \\
P^2 = M^2\\
P_h^2 = M_h^2 \\
2\,P\cdot q = \frac{Q^2}{\xbj} \\
2\,P_h\cdot q = -z_h\,Q^2
\end{array} \right\}
\longleftrightarrow \left\{
\begin{array}{l}
P_h = \frac{z_h\,Q}{\sqrt{2}}\,n_-
+ \frac{M_h^2}{z_h\,Q\sqrt{2}}\,n_+
\\ \mbox{} \\
q =\ \frac{Q}{\sqrt{2}}\,n_- \ - \ \frac{Q}{\sqrt{2}}\,n_+\ +\ q_T
\\ \mbox{} \\
P = \frac{\xbj M^2}{Q\sqrt{2}}\,n_-
+ \frac{Q}{\xbj \sqrt{2}}\,n_+
\end{array}
\right.
\]
Note that this works for socalled current fragmentation,
in which case the produced hadron is {\em hard} with respect
to the target momentum, i.e. $P\cdot P_h \sim Q^2$.
The minus component $p^-$ is irrelevant in the lower
soft part, while the plus component $k^+$ is irrelevant in the upper
soft part. Note that after the choice of $P$ and $P_h$ one can no
longer omit a transverse component in the other vector, in this
case the momentum transfer $q$. This is
precisely the vector $q_T$ introduced earlier in the discussion
of the structure functions for 1-particle inclusive leptoproduction.
One immediately sees that one can no longer simply integrate
over the transverse component of the quark momentum, defined in
Eq.~\ref{quarkmom}.

At this point it turns out that the most convenient way to describe
the spin vector of the target is via an expansion of the form
\bea
S^\mu =
-\lambda\,\frac{M\xbj}{Q\sqrt{2}}\,n_-
+ \lambda\,\frac{Q}{M\xbj\sqrt{2}}\,n_+ + S_\st.
\eea
One has up to ${\cal O}(1/Q^2)$ corrections
$\lambda \approx M\,(S\cdot q)/(P\cdot q)$ and
$S_\st \approx S_\perp$. For a pure state one has
$\lambda^2 + \bmm S_\st^2 = 1$,
in general this quantity being less or equal than one.

\begin{figure}[b]
\leavevmode
\begin{center}
\begin{minipage}{9.0cm}
\epsfig{file=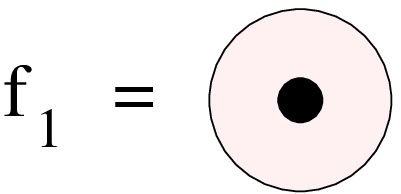,width=1.8cm}
\hspace{2.5 cm}
\epsfig{file=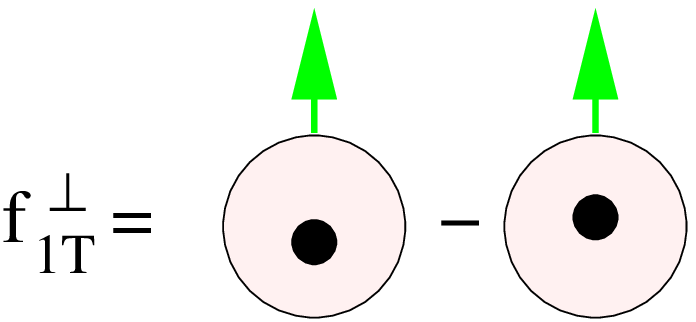,width=3.1cm}
\\[0.2cm]
\epsfig{file=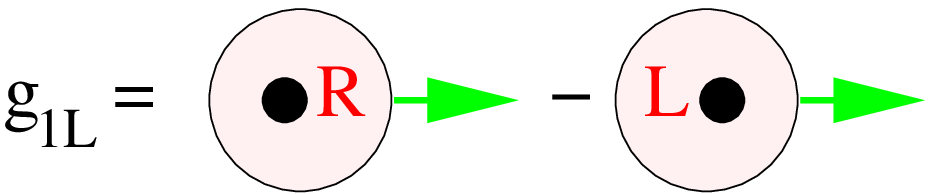,width=4.2cm}
\hspace{0.5 cm}
\epsfig{file=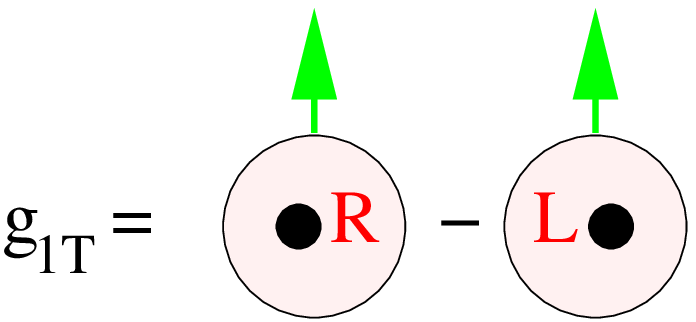,width=3.1cm}
\\[0.2cm]
\epsfig{file=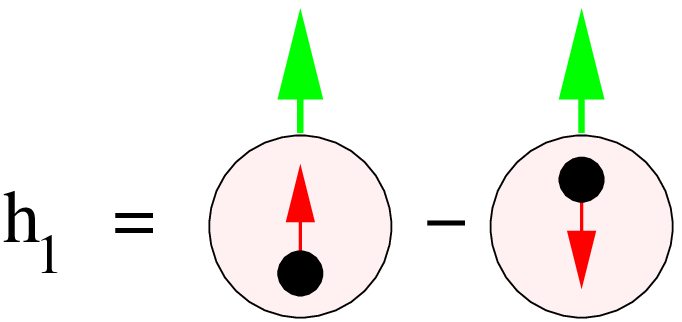,width=3.1cm}
\hspace{1.3cm}
\epsfig{file=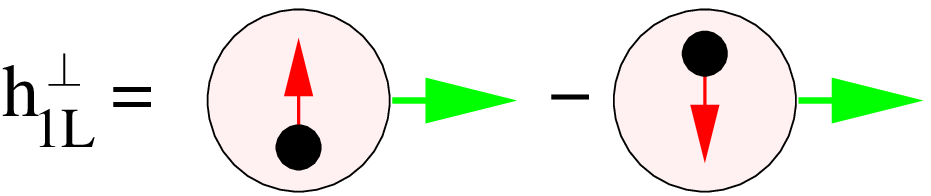,width=4.2cm}
\\[0.6cm]
\epsfig{file=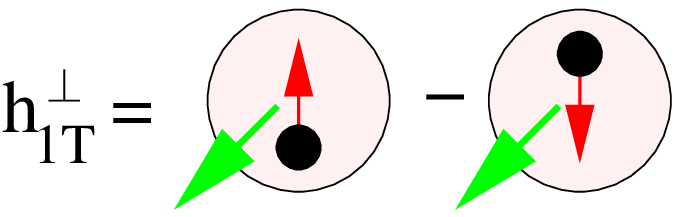,width=3.1cm}
\hspace{1.3 cm}
\epsfig{file=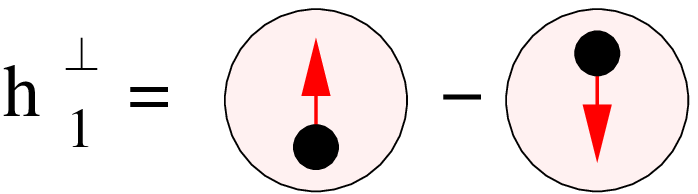,width=3.1cm}
\end{minipage}
\end{center}
\caption{
Interpretation of the functions in the leading
Dirac traces of $\Phi$.
\label{fig3}
}
\end{figure}

The soft part to look at is
\be
\Phi(x,\bmm p_T) =
\left. \int \frac{d\xi^-d^2\bmm \xi_T}{(2\pi)^3}\ e^{ip\cdot \xi}
\,\langle P,S\vert \overline \psi(0) \psi(\xi)
\vert P,S\rangle \right|_{\xi^+ = 0}.
\ee
For the leading order results, it is parametrized as
\be
\Phi(x,\bmm p_\st)  =
\Phi_O(x,\bmm p_\st) + \Phi_L(x,\bmm p_\st) + \Phi_T(x,\bmm p_\st),
\ee
with the parts involving unpolarized
targets (O), longitudinally polarized targets (L) and transversely
polarized targets (T) up to parts proportional to $M/P^+$ given by
\bea
\Phi_O(x,\bmm p_\st) & = &
\frac{1}{2} \Biggl\{
f_1(x,\bmm p_\st)\,\nslash_+
+ h_1^\perp(x,\bmm p_\st)\,\frac{i\,[\pslash_\st,\nslash_+]}{2M}
\Biggr\}
\\
\Phi_L(x,\bmm p_\st) & = &
\frac{1}{2} \Biggl\{
\lambda\,g_{1L}(x,\bmm p_\st)\,\gamma_5\,\nslash_+
+ \lambda\,h_{1L}^\perp(x,\bmm p_\st)
\frac{\gamma_5\,[\pslash_\st,\nslash_+]}{2M}
\Biggr\}
\\
\Phi_T(x,\bmm p_\st)  & = &
\frac{1}{2} \Biggl\{
f_{1T}^\perp(x,\bmm p_\st)\, \frac{\epsilon_{\mu \nu \rho \sigma}
\gamma^\mu n_+^\nu p_\st^\rho S_\st^\sigma}{M}
\nonumber \\ & &\mbox{}
+ \frac{\bmm p_\st\cdot\bmm S_\st}{M}\,g_{1T}(x,\bmm p_\st)
\,\gamma_5\,\nslash_+
+ h_{1T}(x,\bmm p_\st)\,\frac{\gamma_5\,[\Sslash_\st,\nslash_+]}{2}
\nonumber \\ & &\mbox{}
+ \frac{\bmm p_\st\cdot\bmm S_\st}{M}\,h_{1T}^\perp(x,\bmm p_\st)\,
\frac{\gamma_5\,[\pslash_\st,\nslash_+]}{2M}
\Biggr\}.
\eea
All functions appearing here have a natural interpretation
as densities. This is seen as discussed before for the
$\bmm p_\st$-integrated functions. Now it includes densities such as
the density of
longitudinally polarized quarks in a transversely polarized nucleon
($g_{1T}$) and the density of transversely polarized quarks in
a longitudinally polarized nucleon ($h_{1L}^\perp$). The interpretation
of all functions is illustrated in Fig.~\ref{fig3}.

Several functions vanish from the soft part upon integration
over $p_\st$. Actually we will find that particularly interesting
functions survive when one integrates over $\bmm p_\st$ weighting
with $p_\st^\alpha$, e.g.
\bea
\Phi_\partial^\alpha (x) \equiv
\int d^2 p_\st\,\frac{p_\st^\alpha}{M} \,\Phi(x,\bmm p_\st)
& = & \frac{1}{2}\,\Biggl\{
-g_{1T}^{(1)}(x)\,S_\st^\alpha\,\slash n_+\gamma_5
-\lambda\,h_{1L}^{\perp (1)}(x)
\,\frac{[\gamma^\alpha,\slash n_+]\gamma_5}{2}
\nonumber \\ &&\mbox{}
-{f_{1T}^{\perp (1)}}
\,\epsilon^{\alpha}_{\ \ \mu\nu\rho}\gamma^\mu n_-^\nu {S_\st^\rho}
- {h_1^{\perp (1)}}
\,\frac{i[\gamma^\alpha, \nslash_+]}{2}\Biggr\},
\label{Phid}
\eea
where we define $\bmm p_\st^2/2M^2$-moments, referred to as transverse moments,
as
\be
g_{1T}^{(n)}(x) = \int d^2p_\st\ \left(\frac{\bmm p_\st^2}{2M^2}\right)^n
\,g_{1T}(x,\bmm p_\st),
\ee
and similarly the other functions. The use of these moments also turns out
to be important when one tries to incorporate perturbative effects that
affect the $p_\st$-shape~\cite{Boer-00}.
To interpret all functions it is simplest to again write down the matrix
$(\Phi\slash n_-)^T$, which becomes
\be
(\Phi\slash n_-)^T =
\left\lgroup \begin{array}{cccc}
f_1 + g_{1} &
\frac{\vert p_\st\vert}{M}\,e^{i\phi}\,g_{1T}&
\frac{\vert p_\st\vert}{M}\,e^{-i\phi}\,h_{1L}^\perp&
2\,h_{1} \\
& & & \\
\frac{\vert p_\st\vert}{M}\,e^{i\phi}\,g_{1T}^\ast&
f_1 - g_{1} &
\frac{\vert p_\st\vert^2}{M^2}e^{-2i\phi}\,h_{1T}^\perp &
-\frac{\vert
p_\st\vert}{M}\,e^{-i\phi}\,h_{1L}^{\perp\ast}\\ & & &
\\ \frac{\vert p_\st\vert}{M}\,e^{i\phi}\,h_{1L}^{\perp\ast}&
\frac{\vert p_\st\vert^2}{M^2}e^{2i\phi}\,h_{1T}^\perp &
f_1 - g_{1} &
-\frac{\vert p_\st\vert}{M}\,e^{i\phi}\,g_{1T}^\ast \\
& & & \\
 2\,h_{1}&
-\frac{\vert p_\st\vert}{M}\,e^{i\phi}\,h_{1L}^\perp&
-\frac{\vert p_\st\vert}{M}\,e^{-i\phi}\,g_{1T} &
f_1 + g_{1}
\end{array}\right\rgroup ,
\ee
to be compared with Eq.~\ref{prod1}. We have omitted here the T-odd functions
$f_{1T}^\perp$ and $h_1^\perp$ appearing as imaginary parts of $g_{1T}^{\mbox{}}$
and $h_{1L}^\perp$, respectively.
The functions
$h_1^\perp$ and $f_{1T}^\perp$ are T-odd, vanishing if T-reversal
invariance can be applied to the matrix element. For $p_\st$-dependent
correlation functions, matrix elements involving gluon fields at
infinity (gluonic poles~\cite{bmt}) can for instance prevent
application of T-reversal invariance.
The functions describe the possible
appearance of unpolarized quarks in a transversely polarized nucleon
($f_{1T}^\perp$) or transversely polarized quarks in an unpolarized
hadron ($h_1^\perp$) and lead to single-spin asymmetries in various
processes~\cite{Sivers90,Anselmino95}.
The interpretation of all these functions is also
illustrated in Fig.~\ref{fig3}.
Of course just integrating $\Phi(x,p_\st)$ over $p_\st$ gives the
result used in inclusive scattering with $f_1(x)$ = $\int d^2p_\st
\ f_1(x,p_\st)$, $g_1(x)$ = $g_{1L}(x)$ and $h_1(x)$ = $h_{1T}(x)
+ h_{1T}^{\perp (1)}(x)$. We note that the function $h_{1T}^{\perp (2)}$
appears after weighting with $p_\st^\alpha p_\st^\beta$.

\subsection{Bounds}
In analogy to the Soffer bound derived from the production matrix in
Eq.~\ref{prod1} one easily derives a number of new bounds from the full
matrix, such as
\bea
&&f_1(x,\bm p_\st^2) \ge 0 \, ,
\\
&&\vert g_{1}(x,\bm p_\st^2)\vert \le f_1(x,\bm p_\st^2) \,.
\eea
obtained from one-dimensional subspaces
and
\bea
&& \vert h_1 \vert \le
\frac{1}{2}\left( f_1 + g_{1}\right)
\le f_1,
\label{Soffer}\\
&&
\vert h_{1T}^{\perp(1)}\vert \le
\frac{1}{2}\left( f_1 - g_{1}\right)
\le f_1,
\\
&& \vert g_{1T}^{(1)}\vert^2
\le \frac{\bm p_\st^2}{4M^2}
\left( f_1 + g_{1}\right)
\left( f_1 - g_{1}\right)
\le \frac{\bm p_\st^2}{4M^2}\,f_1^2,
\\
&& \vert h_{1L}^{\perp (1)}\vert^2
\le \frac{\bm p_\st^2}{4M^2}
\left( f_1 + g_{1}\right)
\left( f_1 - g_{1}\right)
\le \frac{\bm p_\st^2}{4M^2}\,f_1^2,
\eea
obtained from two-dimensional subspaces.
Here we have used the notation $g_{1T}^{(1)}(x,\bm p_\st^2) \equiv
(\bm p_\st^2/4M^2)\,g_{1T}(x,\bm p_\st^2)$ also for $p_\st$-dependent
functions.
These bounds and their further refinements have been discussed in detail
in Ref.~\cite{BBHM}. There are straightforward extensions of transverse
momentum dependent distribution and fragmentation functions for spin 1
hadrons~\cite{bacchetta} and gluons in spin 1/2 hadrons~\cite{rodrigues}.

\subsection{The relevant fragmentation functions}

Just as for the distribution functions one can perform an analysis of
the soft part describing the quark fragmentation.
One needs~\cite{CS82}
\be
\Delta_{ij}(z,\btt k) =
\left. \sum_X \int \frac{d\xi^+d^2\bmm \xi_\st}{(2\pi)^3} \,
e^{ik\cdot \xi} \,Tr  \langle 0 \vert \psi_i (\xi) \vert P_h,X\rangle
\langle P_h,X\vert\overline \psi_j(0) \vert 0 \rangle
\right|_{\xi^- = 0}.
\ee
For the production of unpolarized hadrons $h$ in hard processes one needs
to leading order in $1/Q$ the correlation function,
\be
\Delta_O(z,\bmm k_\st) =
z\,D_1(z,\bmm k^\prime_\st)\,\nslash_-
+ z\,H_1^\perp(z,\bmm k^\prime_\st)\,\frac{i\,[\kslash_\st,\nslash_-]}{2M_h}
+ {\cal O}\left(\frac{M_h}{P_h^-}\right).
\ee
when we limit ourselves to an unpolarized or spin 0 final state hadron.
The arguments of the fragmentation functions $D_1$ and $H_1^\perp$ are
$z$ = $P_h^-/k^-$ and $\bmm k^\prime_\st$ = $-z\btt k$. The first
is the (lightcone) momentum fraction of the produced hadron, the second
is the transverse momentum of the produced hadron with respect to the quark.
The fragmentation function $D_1$ is the equivalent of the distribution
function $f_1$. It can be interpreted as the probability of finding a
hadron $h$ in a quark.
The function $H_1^\perp$, interpretable as the difference in
production probabilities of unpolarized hadrons from a transversely
polarized quark depending on transverse momentum, is allowed
because of the non-applicability of time reversal
invariance~\cite{Collins93}.
This is natural for the fragmentation functions~\cite{RKR71,JJ93}
because of the  appearance of out-states
$\vert P_h, X\rangle$ in the definition of $\Delta$, in contrast
to the plane wave states appearing in $\Phi$.
After $\btt k$-averaging one is left with the functions
$D_1(z)$ and the $\bmm k_\st^2/2M^2$-weighted result $H_1^{\perp (1)}(z)$.

\subsection{The semi-inclusive cross section}

After the analysis of the soft parts, the next step is to find
out how one obtains the information on the various correlation functions
from experiments, in this particular case in lepton-hadron scattering
via one-photon exchange as discussed before.
To get the leading order result for semi-inclusive scattering it is
sufficient to compute the diagram in Fig.~\ref{fig2} (right)
by using QCD and QED Feynman rules in the hard part and the
matrix elements $\Phi$ and $\Delta$ for the soft parts, parametrized in
terms of distribution and fragmentation functions. The most
well-known results for leptoproduction are:
\bea
&&\frac{d\sigma_{OO}}{d\xbj\,dy\,dz_h}
= \frac{2\pi \alpha^2\,s}{Q^4}\,\sum_{a,\bar a} e_a^2
\left\lgroup 1 + (1-y)^2\right\rgroup \xbj {f^a_1}(\xbj)\,{ D^a_1}(z_h)
\\ && \frac{d\sigma_{LL}}{d\xbj\,dy\,dz_h}
= \frac{2\pi \alpha^2\,s}{Q^4}\,{\lambda_e\,\lambda}
\,\sum_{a,\bar a} e_a^2\  y (2-y)\  \xbj {g^a_1}(\xbj)\,{D^a_1}(z_h)
\eea
The indices attached to the cross section refer to polarization
of lepton (O is unpolarized, L is longitudinally polarized) and
hadron (O is unpolarized, L is longitudinally polarized, T is
transversely polarized). Note that the result is a weighted sum
over quarks and antiquarks involving the charge $e_a$ squared.
Comparing with well-known formal expansions of the cross section in
terms of structure functions one can simply identify these.
For instance the above result for unpolarized scattering (OO)
shows that after averaging over azimuthal angles,
only one structure function survives if we work at order $\alpha_s^0$
and at leading order in $1/Q$.

As we have seen, in 1-particle inclusive unpolarized leptoproduction
in principle four structure functions appear, two of them containing azimuthal
dependence of the form $\cos (\phi_h^\ell)$ and $\cos (2\phi_h^\ell)$.
The first one only appears at order $1/Q$~\cite{LM94}, the second one
even at leading order but only in the case of the existence
of nonvanishing T-odd distribution functions. To be specific if we
define weighted cross section such as
\be
\int d^2\bmm q_{T}\,\frac{Q_{T}^2}{MM_h} \,\cos(2\phi_h^\ell)
\,\frac{d\sigma_{{OO}}}{d\xbj\,dy\,dz_h\,d^2\bmm q_{T}}
\equiv
\left< \frac{Q_{T}^2}{MM_h} \,\cos(2\phi_h^\ell)\right>_{OO}
\ee
we obtain the following asymmetry,.
\be
\left<
\frac{Q_{T}^2}{MM_h} \,\cos(2\phi_h^\ell)\right>_{OO}
= \frac{16\pi \alpha^2\,s}{Q^4}
\,(1-y)\,\sum_{a,\bar a} e_a^2
\,\xbj\,{h_{1}^{\perp(1)a}}(\xbj) H_1^{\perp (1)a}.
\ee
In lepton-hadron scattering this asymmetry requires T-odd distribution
functions and therefore most likely is absent or very small. In
$e^+e^-$ annihilation~\cite{BJM}, however, a $\cos 2\phi$ asymmetry between
produced particles (e.g. pions) in opposite jets involves two
very likely nonvanishing fragmentation functions $H_1^\perp$ and
$\overline H_1^\perp$. Indications for the presence of these
fragmentation functions have been found in LEP data~\cite{Efremov}.

Also in leptoproduction indications for azimuthal asymmetries have
been found~\cite{HERMES,SMC,HERMES2}, which in case of single spin
asymmetries point towards a T-odd fragmentation function.
For polarized targets, several azimuthal asymmetries arise already
at leading order. For example the following possibilities were
investigated in Refs~\cite{KM96,Collins93,Kotzinian95,TM95b}.
\bea
&&
\left< \frac{Q_\st}
{M} \,\cos(\phi_h^\ell-\phi_S^\ell)\right>_{LT} =
\frac{2\pi \alpha^2\,s}{Q^4}\,{\lambda_e\,\vert \btt S \vert}
\,y(2-y)\sum_{a,\bar a} e_a^2
\,\xbj\,{g_{1T}^{(1)a}}(\xbj) {D^a_1}(z_h),
\nonumber \\ && \label{asbas}
\\
&&
\left< \frac{Q_\st^2}{MM_h}
\,\sin(2\phi_h^\ell)\right>_{OL} =
-\frac{4\pi \alpha^2\,s}{Q^4}\,{\lambda}
\,(1-y)\sum_{a,\bar a} e_a^2
\,\xbj\,{h_{1L}^{\perp(1)a}}(\xbj) {H_1^{\perp(1)a}}(z_h),
\nonumber \\ && \label{as2}
\\
&&
\left< \frac{Q_\st}{M_h}
\,\sin(\phi_h^\ell+\phi_S^\ell)\right>_{OT} =
\frac{4\pi \alpha^2\,s}{Q^4}\,{\vert \btt S \vert}
\,(1-y)\sum_{a,\bar a} e_a^2
\,\xbj\,{h_1^a}(\xbj) {H_1^{\perp(1)a}}(z_h).
\nonumber \\ && \label{finalstate}
\eea
The latter two are single spin asymmetries involving the fragmentation
function $H_1^{\perp (1)}$. The last one was the asymmetry proposed by
Collins~\cite{Collins93} as a way to access the transverse spin
distribution function $h_1$ in pion production.
Note, however, that in using the
azimuthal dependence one needs to be very careful. For instance, besides
the $<\sin (\phi_h^\ell + \phi_S^\ell)>_{OT}$, one also finds at leading
order a $<\sin (3\phi_h^\ell - \phi_S^\ell)>_{OT}$ asymmetry which is
proportional to $h_{1T}^{\perp (2)}\,H_1^{\perp (1)}$~\cite{TM95b}.

\section{Inclusion of subleading contributions}

\subsection{Subleading inclusive leptoproduction}

If one proceeds up to order $1/Q$ one also needs terms in
the parametrization of the soft part proportional to
$M/P^+$. Limiting ourselves to the $\bmm p_\st$-integrated
correlations one needs
\begin{eqnarray}
\Phi(x) & = &
\frac{1}{2}\,\Biggl\{
f_1(x)\,\nslash_+
+ \lambda\,g_1(x)\, \gamma_5\,\nslash_+
+ h_1(x)\,\frac{\gamma_5\,[\Sslash_\st,\nslash_+]}{2}\Biggr\}
\nonumber \\
& + & \frac{M}{2P^+}\Biggl\{
e(x) + g_T(x)\,\gamma_5\,\Sslash_\st
+ \lambda\,h_L(x)\,\frac{\gamma_5\,[\nslash_+,\nslash_-]}{2} \Biggr\}
\nonumber \\
& + & \frac{M}{2P^+}\Biggl\{
-\lambda\,e_L(x)\,i\gamma_5
- f_T(x)\,\epsilon_\st^{\rho\sigma}\gamma_\rho S_{\st\sigma}
+ h(x)\,\frac{i\,[\nslash_+,\nslash_-]}{2} \Biggr\}.
\end{eqnarray}
The last set of three terms proportional to $M/P^+$ vanish when
time-reversal invariance applies.

Actually in the calculation of the cross section one has to be
careful. Let us use inclusive scattering off a transversely polarized
nucleon (transverse means $\vert \bmm S_\perp\vert = 1$ in
Eq.~\ref{inclspin}) as an example. The hadronic tensor is zero in
leading order in $1/Q$. At order $1/Q$ one obtains from the handbag
diagram a contribution
\be
2M\,W_{A (a)}^{\mu\nu}({q,P,S_\st}) =
i\,\frac{2M}{Q}
\,\hat t_{\mbox{}}^{\,[\mu}\epsilon_\perp^{\nu ]\rho} {S_{\perp\rho}}
\,\left( g_{1T}^{(1)}(\xbj) - \frac{m}{M}\,h_1(\xbj)\right).
\ee
It shows that one must be very careful with the integration over $p_\st$.

\begin{figure}[t]
\begin{center}
\epsfig{file=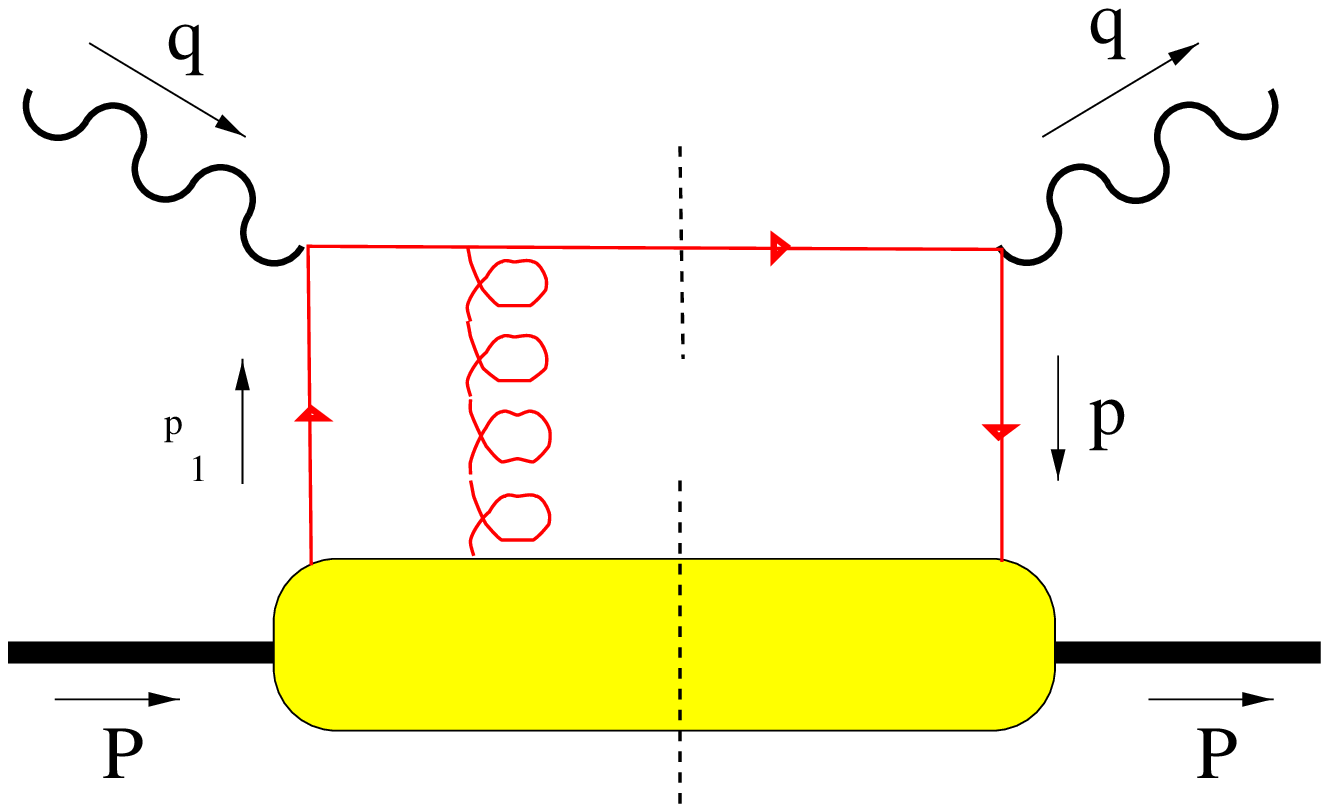,width=4.5cm}
\hspace{1cm}
\epsfig{file=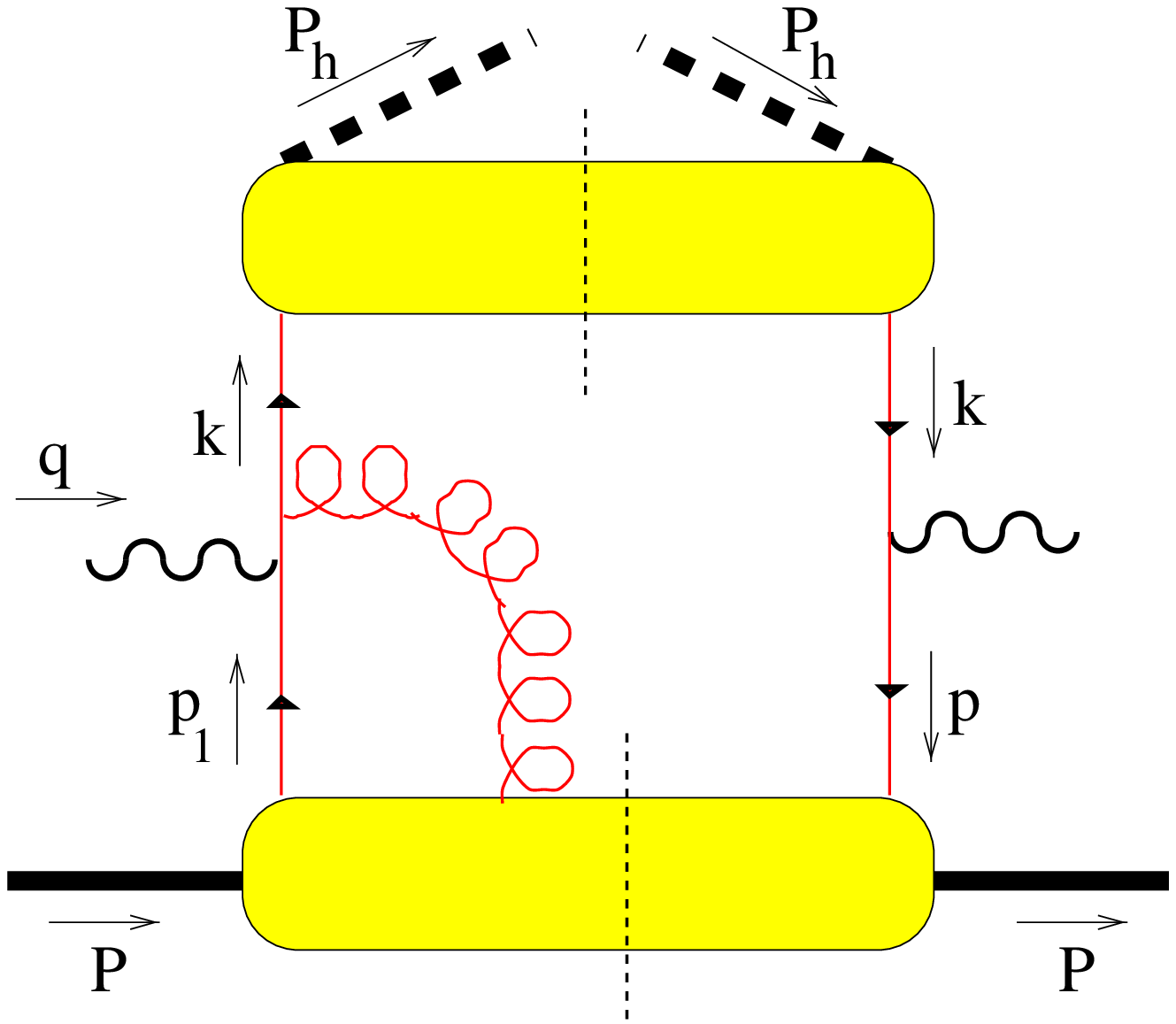,width=4.5cm}
\end{center}
\caption{\label{figdis1}
Examples of gluonic diagrams that must be included at subleading order
in lepton hadron inclusive scattering (left) and in semi-inclusive
scattering (right).}
\end{figure}

There is a second contribution at order $1/Q$ coming from diagrams as the
one shown in Fig.~\ref{figdis1}. For these gluon diagrams one needs
matrix elements
containing $\overline \psi(0)\,gA_\st^\alpha(\eta)\,\psi(\xi)$. At order
$1/Q$ one only needs the matrix element of the bilocal combinations
$\overline \psi(0)\,gA_\st^\alpha(\xi)\,\psi(\xi)$ and
$\overline \psi(0)\,gA_\st^\alpha(0)\,\psi(\xi)$
These soft parts have a structure quite similar to $\Phi_\partial^\alpha$
and are parametrized as
\bea
\Phi_A^\alpha (x) &= &
\frac{M}{2}\,\Biggl\{
-x\,\tilde g_{T}(x)\,S_\st^\alpha\,\slash n_+\gamma_5
-\lambda\,x\,\tilde h_{L}(x)
\,\frac{[\gamma^\alpha,\slash n_+]\gamma_5}{2}
\nonumber \\
&&\quad \mbox{}
-x\,\tilde f_T(x)
\,\epsilon^{\alpha}_{\ \ \mu\nu\rho}\gamma^\mu n_-^\nu {S_\st^\rho}
- x\,\tilde h(x)
\,\frac{i[\gamma^\alpha, \nslash_+]}{2}\Biggr\}.
\label{phiA}
\eea
This contributes also to $W_A^{\mu\nu}$,
\be
2M\,W_{A (b)}^{\mu\nu}({q,P,S_\st}) =
i\,\frac{2M\xbj}{Q}
\,\hat t_{\mbox{}}^{\,[\mu}\epsilon_\perp^{\nu ]\rho} {S_{\perp\rho}}
\,\tilde g_T(\xbj).
\ee
Using the QCD equations of motion, however, these functions can be related
to the functions appearing in $\Phi$. To be precise one combines $i\partial$ in
$\Phi_\partial$ (see Eq.~\ref{Phid})
and $A_\mu$ in $\Phi_A$ to $\Phi_D$ containing
$iD_\mu = i\partial_\mu + g\,A_\mu$ for which one has via the equations
of motion
\bea
\Phi_D^\alpha (x) &= &
\frac{M}{2}\,\Biggl\{
-\left(x\,g_{T}- \frac{m}{M}\,h_1\right)\,S_\st^\alpha\,\slash n_+\gamma_5
\nonumber \\
&&\quad \mbox{}
-\lambda\left(x\,h_{L}-\frac{m}{M}\,g_1\right)
\,\frac{[\gamma^\alpha,\slash n_+]\gamma_5}{2}
\nonumber \\
&&\quad \mbox{}
-x\,f_T(x)
\,\epsilon^{\alpha}_{\ \ \mu\nu\rho}\gamma^\mu n_-^\nu {S_\st^\rho}
- x\,\tilde h(x)
\,\frac{i[\gamma^\alpha, \nslash_+]}{2}\Biggr\}.
\eea
Hence one obtains
\bea
&&
x\,\tilde g_T = x\,g_T - g_{1T}^{(1)} - \frac{m}{M}\,h_1,
\\
&&
x\,\tilde h_L = x\,h_L - h_{1L}^{\perp (1)} - \frac{m}{M}\,g_1,
\\
&&
x\,\tilde f_T = x\,f_T + f_{1T}^{(1)},
\\
&&
x\,\tilde h = x\,h + 2\,h_{1}^{\perp (1)}.
\eea
and one obtains the full contribution
\be
2M\,W_{A}^{\mu\nu}({q,P,S_\st}) =
i\,\frac{2M\xbj}{Q}
\,\hat t_{\mbox{}}^{\,[\mu}\epsilon_\perp^{\nu ]\rho} {S_{\perp\rho}}
\,g_T(\xbj),
\ee
leading for the structure function ${\it g}_T(\xbj,Q^2)$ defined in
Eq.~\ref{wanti} to the result
\be
{\it g}_T(\xbj,Q^2) = \frac{1}{2}\sum_a e_a^2
\left( g_T^a(\xbj) + g_T^{\bar a}(\xbj)\right).
\ee

{}From Lorentz invariance one obtains, furthermore, some interesting
relations between the subleading functions and the $k_\st$-dependent
leading functions~\cite{BKL84,MT96,BM98}. Just by using the expressions
for the functions in terms of the amplitudes $A_i$ in Eq.~\ref{lorentz}
one finds
\bea
&&g_T  = g_1 + \frac{d}{dx}\,g_{1T}^{(1)},
\label{gTrel}
\\
&&h_L = h_1 - \frac{d}{dx}\,h_{1L}^{\perp (1)},
\label{hLrel}
\\
&&f_T =  - \frac{d}{dx}\,f_{1T}^{\perp (1)},
\\
&&h =  - \frac{d}{dx}\,h_{1}^{\perp (1)}.
\label{rel4}
\eea
As an application, one can eliminate $g_{1T}^{(1)}$ using Eq.~\ref{gTrel}
and obtain (assuming sufficient neat behavior of the functions)
for $g_2 = g_T - g_1$
\bea
g_2(x) & = & -\left[ g_1(x) - \int_x^1 dy \,\frac{g_1(y)}{y} \right]
+ \frac{m}{M} \left[ \frac{h_1(x)}{x}-\int_x^1 dy\, \frac{h_1(y)}{y^2}\right]
\nonumber \\
&& \mbox{}
+ \left[\tilde g_T(x) - \int_x^1 dy\, \frac{\tilde g_T (y)}{y}\right].
\label{wweq}
\eea
One can use this to obtain for each quark flavor $\int dx\,g^a_2(x) = 0$, the
Burkhardt-Cottingham sumrule~\cite{BC}.
Neglecting the interaction-dependent part one
obtains the Wandzura-Wilczek approximation~\cite{WW}
for $g_2$, which in particular
when one neglects the quark mass term provides a simple and often used
estimate for $g_2$. It has become the standard with which experimentalists
compare the results for $g_2$.

Actually the SLAC results for $g_2$ can also be used to estimate the
function $g_{1T}^{(1)}$ and the resulting asymmetries, e.g. the
one in Eq.~\ref{asbas}. For this one needs the exact relation
in Eq.~\ref{gTrel}. Results can be found in Refs~\cite{KM96} and
\cite{BM99}.

The full set of
results for the twist-3 functions, including as the first one the above
Wandzura-Wilczek result in Eq.~\ref{wweq} are (omitting quark mass
terms)
\bea
&&
g_T(x) =
\int_x^1 dy\ \frac{g_1(y)}{y}
+ \left[ \tilde g_T(x) - \int_x^1 dy\ \frac{\tilde g_T(y)}{y}\right],
\\ &&
\frac{g_{1T}^{(1)}(x)}{x} =
\int_x^1 dy\ \frac{g_1(y)}{y}
- \int_x^1 dy\ \frac{\tilde g_T(y)}{y},
\\[0.5cm] &&
h_L(x) =
2x\int_x^1 dy\ \frac{h_1(y)}{y^2}
+ \left[ \tilde h_L(x) - 2x\int_x^1 dy\ \frac{\tilde h_L(y)}{y^2}\right],
\\ &&
\frac{h_{1L}^{\perp(1)}(x)}{x^2} =
-\int_x^1 dy\ \frac{h_1(y)}{y^2}
+ \int_x^1 dy\ \frac{\tilde h_L(y)}{y^2},
\\[0.5cm] &&
f_T(x) =
\left[ \tilde f_T(x) - \int_x^1 dy\ \frac{\tilde f_T(y)}{y}\right] ,
\\ &&
\frac{f_{1T}^{\perp (1)}(x)}{x} =
\int_x^1 dy\ \frac{\tilde f_T(y)}{y},
\\[0.5cm] &&
h(x) =
\left[ \tilde h(x) - 2x\int_x^1 dy\ \frac{\tilde h(y)}{y^2}\right],
\\ &&
\frac{h_{1}^{\perp(1)}(x)}{x^2} =
\int_x^1 dy\ \frac{\tilde h(y)}{y^2} .
\eea
The Wandzura-Wilczek result has
a complete analogue in the relation for $h_L$ discussed in Ref.~\cite{JJ92}.
In slightly different form the result for $g_{1T}^{(1)}$ has been discussed
in Ref.~\cite{BKL84}.
Actually, we need not consider the T-odd functions separately. They can be
simply considered as imaginary parts of other functions, when we allow complex
functions. In particular one can expand the correlation functions into
matrices in Dirac space~\cite{BBHM} to show that the relevant combinations are
$(g_{1T} - i\,f_{1T}^{\perp})$ which we can treat together as one complex
function $g_{1T}$. Similarly we can use complex functions
$(h_{1L}^\perp + i\,h_1^\perp)$ $\rightarrow$ $h_{1L}^\perp$,
$(g_T + i\,f_T)$ $\rightarrow$ $g_T$,
$(h_L + i\,h)$ $\rightarrow$ $h_L$,
$(e + i\,e_L)$ $\rightarrow$ $e$. The functions $f_1$, $g_1$ and $h_1$ remain
real, they don't have T-odd partners.

\subsection{Subleading 1-particle inclusive leptoproduction}

Also for the transverse momentum dependent functions dependent distribution
and fragmentation functions one can proceed to subleading order
\cite{MT96}, but very likely one will also find competing $\alpha_s$
corrections contributing to the same observables.
We will not discuss these functions here.

In semi-inclusive cross sections one also needs
fragmentation functions, for which similar relations exist, e.g. the
relation in Eq.~\ref{rel4} for distribution functions
has an analog for fragmentation functions, relating
$H_1^{\perp (1)}$ (appearing in Eqs~\ref{as2} and \ref{finalstate})
and an at subleading order appearing function $H(z)$,
\be
\frac{H(z)}{z} = z^2\,\frac{d}{dz} \left(\frac{H_1^{\perp (1)}}{z}\right).
\label{frag1}
\ee
The full relations can be found in \cite{HBM}.

An interesting subleading asymmetry involving $H_1^\perp$ is
a $\sin(\phi_h^\ell)$ single spin asymmetry appearing as the structure
functions ${\cal H}_{LT}^\prime$ in Eq.~\ref{sidiswanti} for a polarized
lepton but unpolarized target~\cite{LM94},
\bea
&&\left< \frac{Q_{T}} {M} \,\sin(\phi_h^\ell) \right>_{LO} =
\frac{4\pi \alpha^2\,s}{Q^4}\,{\lambda_e}
\,y\sqrt{1-y}
\,\frac{2M}{Q}\,\xbj^2 {\tilde e^a}(\xbj)\,{H_1^{\perp (1)a}}(z_h)
\label{as1}
\eea
where $\tilde e^a(x) = e^a(x) - (m_a/M)\,(f_1^a(x)/x)$.
This cross section involves, besides the
time-reversal odd fragmentation function $H_1^\perp$,
the distribution function $e$.
The tilde function that appear in the cross sections is in fact
the socalled interaction dependent part of the twist three
functions. It would vanish in any naive parton model calculation in
which cross sections are obtained by folding electron-parton cross
sections with parton densities. Considering the relation for $\tilde e$
one can state it as $x\,e(x)$ = $(m/M)\,f_1(x)$ in the absence of
quark-quark-gluon correlations. The inclusion of the latter also
requires diagrams dressed with gluons as shown in Fig.~\ref{figdis1}.

\section{Color gauge invariance}

We have sofar neglected two problems. The first problem is that
the correlation function $\Phi$ discussed in previous sections involve
two quark fields at different space-time points and hence are not
color gauge invariant.
The second problem comes from the gluonic diagrams similar as the
ones we have discussed in the previous section (see Fig.~\ref{figdis1})
We note that diagrams involving matrix elements with longitudinal ($A^+$)
gluon fields,
\[
\overline \psi_j(0)\,gA^+(\eta)\,\psi_i(\xi)
\]
do not lead to any suppression. The reason is that because of the
$+$-index in the gluon field
the matrix element is proportional to $P^+$, $p^+$ or
$M\,S^+$ rather than the proportionality to $M\,S_\st^\alpha$ or
$p_\st^\alpha$ that we have seen in Eq.~\ref{phiA} for a gluonic
matrix element with transverse gluons.

A straightforward calculation, however, shows that the gluonic diagrams with
one or more longitudinal gluons involve matrix elements (soft parts)
of operators $\overline \psi \psi$,
$\overline \psi\,A^+\,\psi$, $\overline \psi\,A^+A^+\,\psi$, etc.
that can be resummed into a correlation function
\be
\Phi_{ij}(x) =
\left. \int \frac{d\xi^-}{2\pi}\ e^{ip\cdot \xi}
\,\langle P,S\vert \overline \psi_j(0)\,{\cal U}(0,\xi)\,\psi_i(\xi)
\vert P,S\rangle \right|_{\xi^+ = \xi_\st = 0},
\ee
where ${\cal U}$ is a gauge link operator
\be
{\cal U}(0,\xi)
= {\cal P}\exp\left(-i\int_0^{\xi^-} d\zeta^-\,A^+(\zeta)\right)
\ee
(path-ordered exponential with path along $-$-direction).
Et voila, the unsuppressed gluonic diagrams combine into
a color gauge invariant correlation function.
We note that at the level of operators, one expands
\be
\overline \psi(0)\psi(\xi) =
\sum_n \frac{\xi^{\mu_1}\ldots \xi^{\mu_n}}{n!}\,
\overline \psi(0)\partial_{\mu_1}\ldots\partial_{\mu_n}\psi(0),
\ee
in a set of local operators, but only
the expansion of the nonlocal combination with a gauge link
\be
\overline \psi(0)\psi(\xi) =
\sum_n \frac{\xi^{\mu_1}\ldots \xi^{\mu_n}}{n!}\,
\overline \psi(0)D_{\mu_1}\ldots D_{\mu_n}\psi(0),
\ee
is an expansion in terms of local gauge invariant operators.
The latter operators are precisely the local (quark) operators
that appear in the operator product expansion applied to
inclusive deep inelastic scattering.

For the $p_\st$-dependent functions, one finds that inclusion of
$A^+$ gluonic diagrams leads to a color gauge invariant matrix
element with links running via $\xi^= = \pm \infty$~\cite{BM00}.
For instance
in lepton-hadron scattering one finds
\be
\Phi(x,\bmm p_T) =
\left. \int \frac{d\xi^-d^2\bmm \xi_T}{(2\pi)^3}\ e^{ip\cdot \xi}
\,\langle P,S\vert \overline \psi(0)\,{\cal U}(0,\infty)
\,{\cal U}(\infty,\xi)\,\psi(\xi)
\vert P,S\rangle \right|_{\xi^+ = 0},
\ee
where the gauge links are at constant $\xi_\st$.
One can multiply this correlator with $p_\st^\alpha$ and make this
into a derivative $\partial_\alpha$. Including the links one finds
the color gauge invariant result
\bea
&&p_\st^\alpha\,\Phi_{ij}(x,\bmm p_\st) =
(\Phi_\partial^{\alpha})_{ij}(x,\bmm p_\st) \nonumber \nonumber \\ &&
\quad =
\int \frac{d\xi^-\,d^2\bmm \xi_\st}{(2\pi)^3}\ e^{ip\cdot \xi}
\biggl\{ \langle P,S\vert \overline \psi_j(0)
\,{\cal U}(0,\infty)\, iD_\st^\alpha\psi_i(\xi) \vert P,S\rangle
\biggr|_{\xi^+=0}
\nonumber \\ && \mbox{}\hspace{2cm}
- \langle P,S\vert \overline \psi_j(0)\,{\cal U}(0,\infty)
\int_{\infty}^{\xi^-}d\eta^- \,{\cal U}(\infty,\eta)
\nonumber \\ && \mbox{}\hspace{3cm}\times g\,G^{+\alpha}(\eta)
\,{\cal U}(\eta,\xi)\,\psi_i(\xi) \vert P,S\rangle \biggr|_{\xi^+=0}\biggr\},
\eea
which gives after integration over $p_\st$ the expected result
$\Phi_\partial^\alpha(x) = \Phi_D^\alpha(x) - \Phi_A^\alpha(x)$. Note
that in $A^+ = 0$ gauge all the gauge links disappear, while one
has $G^{+\alpha} = \partial^+A^\alpha$, but there presence is essential
to perform the above differentiations.

\section{Evolution}

\subsection{Evolution and $p_\st$-dependence}

\begin{figure}[t]
\begin{center}
\epsfig{file=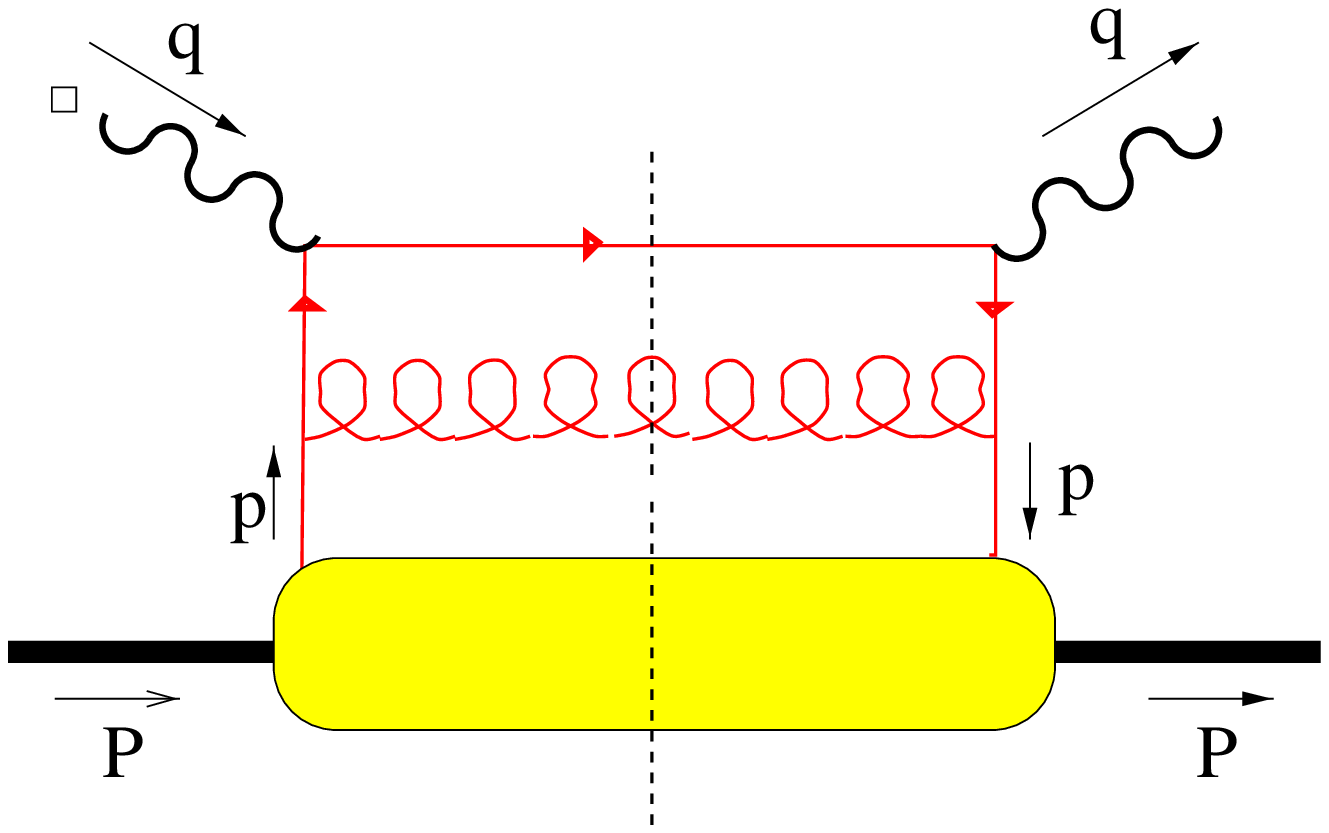,width=4.5cm}
\hspace{1cm}
\epsfig{file=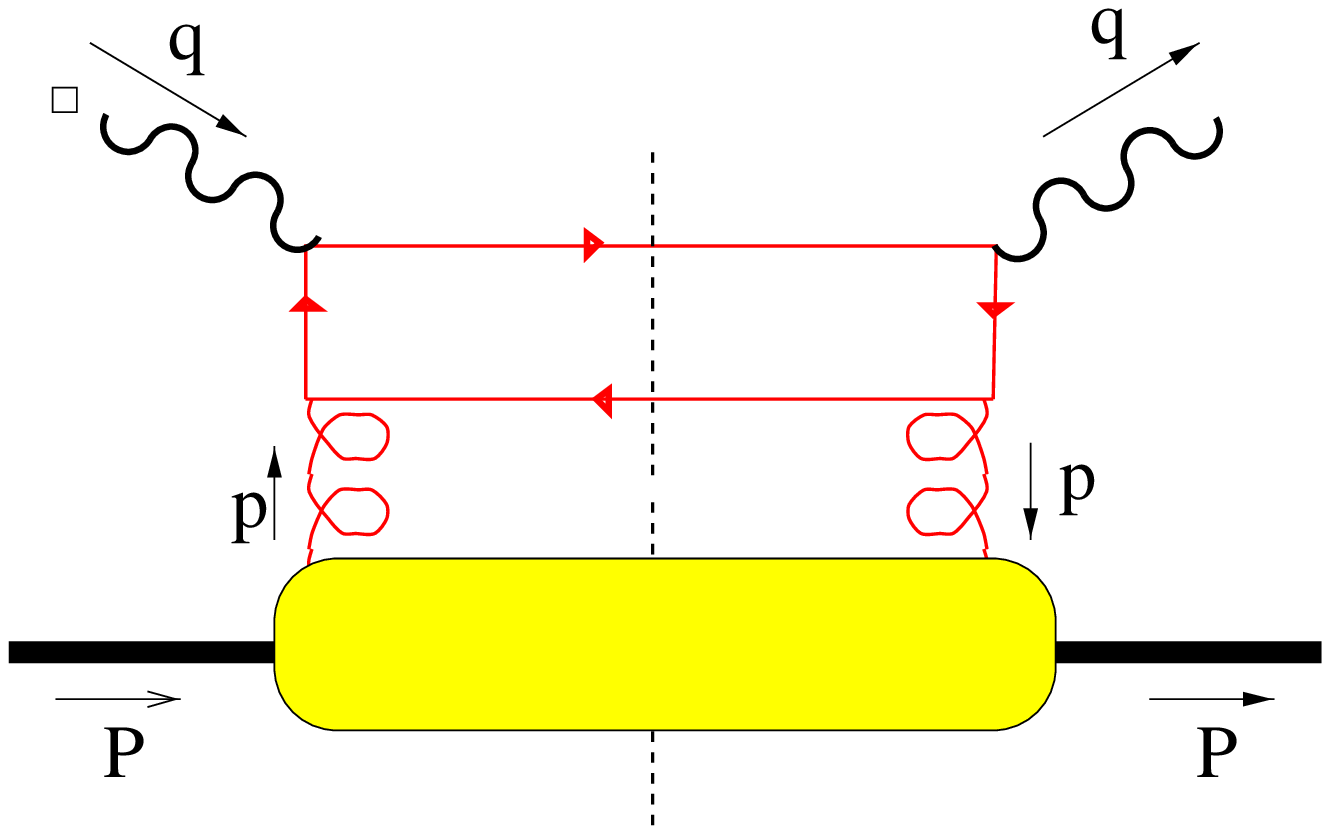,width=4.5cm}
\end{center}
\caption{\label{figdis2}
Ladder diagrams used to calculate the asymptotic behavior of the
correlation functions.}
\end{figure}
The explicit treatment of transverse momenta provides also a transparent
way to include the evolution equations for quark distribution and
fragmentation functions. Remember that we have assumed that soft parts
vanish sufficiently fast as a function of the invariants $p\cdot P$ and
$p^2$, which at constant $x$ implies a sufficiently fast vanishing as
a function of $\bmm p_\st^2$. This simply turns out not to be true.
Assuming that the result for $\bmm p_\st^2 \ge \mu^2$ is given by
the diagram shown in Fig.~\ref{figdis2} one finds
\bea
f_1(x,\bmm p_\st^2) & = & \theta (\mu^2-\bmm p_\st^2)\,f_1(x,\bmm p_\st^2)
\nonumber \\ & + &
\theta(\bmm p_\st^2-\mu^2)
\,\frac{1}{\pi\,\bmm p_\st^2}
\,\frac{\alpha_s(\mu^2)}{2\pi}
\int_x^1 \frac{dy}{y}\,P_{qq}\left(\frac{x}{y}\right)\,f_1(y;\mu^2),
\eea
where $f_1(x;\mu^2) = \pi \int_0^{\mu^2} d\bmm p_\st^2\ f_1(x,\bmm p_\st^2)$
and the splitting function is given by
\be
P_{qq}(z) = C_F\,\left[\frac{1+z^2}{(1-z)_+} + \frac{3}{2}\,\delta(1-z)\right],
\ee
with $\int dz\,f(z)/(1-z)_+ \equiv \int dz\,(f(z)-f(1))/(1-z)$
and the color factor $C_F$ = 4/3 for $SU(3)$.
With the introduction of the scale in $f_1(x;\mu^2)$ one sees that the
scale dependence satisfies
\be
\frac{\partial f_1(x;\mu^2)}{\partial \ln \mu^2}
= \frac{\alpha_s}{2\pi}
\int_x^1 \frac{dy}{y}\,P_{qq}\left(\frac{x}{y}\right)\,f_1(y;\mu^2).
\ee
This is the standard~\cite{Roberts}
nonsinglet evolution equation for the valence quark distribution
function. For the flavor singlet combination of quark distributions or the
sea distributions one also needs to take into account contributions as
shown in Fig.~\ref{figdis2} (right) involving the gluon distribution
functions related to matrix elements with gluon fields $F_{\mu\nu}(\xi)$
but otherwise proceeding along analogous lines.
The $\delta$-function contribution can be explicitly calculated by including
vertex corrections (socalled virtual diagrams), but it is easier to derive
them by requiring that the sum rules for $f_1$ remain valid under evolution,
which requires that $\int_0^1 dz\ P_{qq}(z) = 0$.
\begin{quotation}
\small
Except for logarithmic contributions also finite $\alpha_s$ contributions
show up in deep inelastic scattering~\cite{Roberts}.
For instance in inclusive scattering
one finds that the lowest order result for $F_L$ is of this type,
\begin{eqnarray}
F_L(\xbj,Q^2) & = & \frac{\alpha_s(Q^2)}{4\pi} \Biggl[ C_F \int_{\xbj}^1
\frac{dy}{y}\,\left(\frac{2\xbj}{y}\right)^2 y\,f_1(y;Q^2) \nonumber \\
&&\mbox{} + \left( 2\sum_q e_q^2 \right) \int_{\xbj}^1
\frac{dy}{y} \,\left( \frac{2\xbj}{y} \right)^2 \left( 1 - \frac{\xbj}{y}
\right) \,y\,G(y;Q^2) \Biggr],
\nonumber \\ &&
\end{eqnarray}
the second term involving the gluon distribution function $G(x)$.
\end{quotation}

\subsection{Evolution of transverse moments~\cite{HBM}}

We will present here the evolution of the transverse moments as they are
deduced from known evolution equations for twist-2 functions and for the
interaction dependent tilde functions. For the latter we will only exhibit
the large $N_c$ results, for which the evolution can be written down in
a compact way~\cite{Evol}

As mentioned the evolution of the twist-2 functions and the tilde functions in
known. The twist-2 functions have an autonomous evolution of the form
\be
\frac{d}{d\tau} \,f(x,\tau) = \frac{\alpha_s(\tau)}{2\pi}
\,\int_x^1 \frac{dy}{y} \ P^{[f]}\left(\frac{x}{y}\right)\,f(y,\tau),
\ee
where $\tau$ = $\ln Q^2$ and $P^{[f]}$ are the splitting functions.
The leading order results for the non-singlet twist-2 functions (with the
usual + prescription) \cite{AP,Baldra81} are
\bea
&&
P^{[f_1]}(\beta) = P^{[g_1]}(\beta) = C_F\left[\frac{3}{2}\,\delta(1-\beta) +
\frac{1+\beta^2}{(1-\beta)_+}\right],
\\ &&
P^{[h_1]}(\beta) = C_F\left[\frac{3}{2}\,\delta(1-\beta) +
\frac{2\beta}{(1-\beta)_+}\right].
\eea
In the large $N_c$ limit, also the tilde functions have an autonomous
evolution. In leading order one has
for the interaction-dependent functions \cite{ABH}
\bea
&&
P^{[\tilde f]}(\beta) =  \frac{N_c}{2}\left[
\frac{1}{2}\,\delta(1-\beta) + \frac{2}{(1-\beta)_+} + c\right],
\eea
with $c = -1$ for $\tilde g_T$, $c = -3$ for $\tilde h_L$ and $c=+1$ for
$\tilde e$.

Using these splitting functions and the
relations given in the previous section, we find the evolution of the
transverse moments,
\bea
&&
\frac{d}{d\tau}\,g_{1T}^{(1)}(x,\tau)
= \frac{\alpha_s(\tau)}{4\pi}\,N_c\int_x^1 dy\,\Biggl\{
\left[\frac{1}{2}\,\delta(y-x) + \frac{x^2+xy}{y^2(y-x)_+}\right]
\,g_{1T}^{(1)}(y,\tau)
\nonumber \\ && \hspace{8cm}
+ \frac{x^2}{y^2}\,g_1(y,\tau)\Biggr\} ,
\\ &&
\frac{d}{d\tau}\,h_{1L}^{\perp (1)}(x,\tau)
= \frac{\alpha_s(\tau)}{4\pi}\,N_c\int_x^1 dy\,\Biggl\{
\left[\frac{1}{2}\,\delta(y-x) + \frac{3x^2-xy}{y^2(y-x)_+}\right]
\,h_{1L}^{\perp (1)}(y,\tau)
\nonumber \\ && \hspace{8cm}
-\frac{x}{y}\,h_1(y,\tau)\Biggr\}.
\eea
One can also analyse the fragmentation functions or use some specific
reciprocity relations~\cite{HBM}.
Furthermore, we note that apart from a
$\gamma_5$ matrix the operator structures of the T-odd functions
$f_{1T}^{\perp (1)}$ and $h_1^{\perp (1)}$ are in fact the same
as those of $g_{1T}^{(1)}$ and $h_{1L}^{\perp (1)}$ (or as mentioned before,
they can be considered as the imaginary part of these functions~\cite{BBHM}).
With these ingredients one immediately obtains for the non-singlet functions
the (autonomous) evolution of the T-odd fragmentation functions.
In particular we obtain for
the Collins fragmentation function (at large $N_c$),
\bea
&&\frac{d}{d\tau} \,z H_{1}^{\perp (1)}(z,\tau)
= \frac{\alpha_s}{4\pi}\; N_c\; \int_z^1 dy\, \left[
\frac{1}{2}\,\delta(y-z) + \frac{3y-z}{y(y-z)_+} \right]
\,y H_{1}^{\perp (1)}(y,\tau),
\nonumber \\ &&
\eea
which should prove useful for the comparison of data on Collins
function asymmetries from  different experiments, performed at different
energies.

\section{Concluding remarks}

In these lectures I have discussed aspects of hard scattering processes,
in particular inclusive and 1-particle inclusive lepton-hadron scattering.
The goal is the study of the quark and gluon structure of hadrons.
For example, by considering polarized targets or particle production
one can measure spin and azimuthal asymmetries and use them to obtain
information on specific correlations between
spin and momenta of the partons. The reason why this is a promising
route is the existence of a field theoretical framework that allows
a clean study of the observables as well-defined hadronic matrix elements.

\bigskip
{\small I want to thank the organisers of this school, in particular Prof.
M. Finger, for a pleasant and stimulating meeting.}

\end{document}